# Comprehensive Unified Models of Structural and Reduced Form Models for Defaultable Fixed Income Bonds

(**Part 1:** One factor-model, **Part 2:** Two factors-model)

**Hyong-Chol O** [1], **Song-Yon Kim**, **Dong-Hyok Kim** and **Chol-Hyok Pak**

*Faculty of Mathematics, **Kim Il Sung** University, Pyongyang, D. P. R. Korea*

*Corresponding Authors' E-mail address*: ohyongchol@yahoo.com

This version written on 26 September, 2013

***Abstract*** : Pricing formulae for defaultable corporate bonds with discrete coupons under consideration of the government taxes in the united model of structural and reduced form models are provided. The aim of this paper is to generalize the comprehensive structural model for defaultable fixed income bonds (considered in [1]) into a comprehensive unified model of structural and reduced form models. Here we consider the one factor *model* and the two factor model. In the *one factor model* the bond holders receive the *deterministic* coupon at predetermined coupon dates and the face value (debt) and the coupon at the maturity as well as the effect of government taxes which are paid on the proceeds of an investment in bonds is considered under *constant short rate*. In the *two factor model* the bond holders receive the *stochastic* coupon (*discounted value of that at the maturity*) at predetermined coupon dates and the face value (debt) and the coupon at the maturity as well as the effect of government taxes which are paid on the proceeds of an investment in bonds is considered under *stochastic short rate*. The expected default event occurs when the equity value is not enough to pay coupon or debt at the coupon dates or maturity and unexpected default event can occur at the first jump time of a Poisson process with the given default intensity provided by a step function of time variable. We consider the model and pricing formula for equity value and using it calculate expected default barrier. Then we provide pricing model and formula for defaultable corporate bonds with discrete coupons and consider its duration and the effect of the government taxes.

2010 ***MSC*** : 35C15, 35Q91, 91G20, 91G40, 91G50, 91G80

## 1.   Introduction

The study on defaultable corporate bond is recently one of the most interesting areas of cutting edge in financial mathematics.

As well known, there are two main approaches to pricing defaultable corporate bonds; one is the *structural approach* and the other one is the *reduced form approach*. In the structural method, we think that the default event occurs when the firm value is not enough to repay debt, that is, the firm value reaches a certain lower threshold (*default barrier*) from the above. Such a default can be expected and thus we call it *expected default*. In the reduced-form approach, the default is treated as an unpredictable event governed by a default intensity process. In this case, the default event can occur without any correlation with the firm value and such a default is called *unexpected default*. In the reduced-form approach, if the default probability in time interval $[t, t+\Delta t]$ is $\lambda \Delta t$, then $\lambda$ is called *default intensity* or *hazard rate*. The *third approach* is to unify the structural and reduced form approaches. As for the history of the above three approaches and their advantages and shortcomings, readers can refer to [13] and the introductions of [3, 12]. Combining the elements of the structural approach and reduced-form approach is one of the recent trends.

On the other hand, many models related to coupon approximate actual coupon bearing debts with continuous coupon stream or even zero coupon contracts but such approach has restriction [5].

There has been relatively little work on the most realistic payout structure providing fixed discrete coupons [1]. Geske (1977) is the first study for this problem, where discrete interest payouts prior to maturity were modeled as determinants of default risk [4]. The introduction and the conclusions of [1] includes many useful information about corporate discrete coupon bonds. Recently, Agliardi (2011) generalized the Geske's formula for defaultable coupon bonds, incorporated a stochastic risk free term structure and the effects of bankruptcy cost and government taxes on bond interest and studied the duration of defaultable bonds. Agliardi's approach in [1] to corporate coupon bonds is a kind of structural approaches as shown in its title. In [9], authors studied some general properties of solutions to inhomogeneous Black-Scholes equations with discontinuous maturity payoffs and applied them to a pricing problem of defaultable discrete coupon bond with constant default intensity in a unified model of structural and reduced form models. Unlike [1], the authors of [9] calculated the expected barrier from the bond price.

The aim of this paper is to generalize the comprehensive structural model for defaultable fixed income bonds considered in [1] into a comprehensive unified model of structural and reduced form models. Here we consider the one factor *model* and the two factor model.

In the *one factor model* the bond holders receive the *deterministic* coupon at predetermined coupon dates (like in [1]) and the face value (debt) and the coupon at the maturity. The effect of government taxes which are paid on the proceeds of an investment in bonds is considered under *constant short rate* while Agliardi [1] considered *stochastic* model of short rate such as Vasicek model. The aim of this change is to get analytical pricing formulae. We feel some *difficulty* to get analytical pricing formulae of defaultable discrete (deterministic) coupon bond under unified model of structural and reduced form models and the stochastic models of short rate including Vasicek model.

In the *two factor model* the bond holders receive the *stochastic* coupon (*discounted value of that at the maturity*) at predetermined coupon dates and the face value (debt) and the coupon at the maturity, which is different from [1] and the *aim* of such a change of coupon structure is to get analytical pricing formulae. The effect of government taxes which are paid on the proceeds of an investment in bonds is considered under *stochastic short rate*.

The expected default event occurs when the equity value is not enough to pay coupon or debt at the coupon dates or maturity and unexpected default event can occur at the first jump time of a Poisson process with the given default intensity (provided by a step function of time variable). We consider the model and pricing formula for equity value and using it calculate expected default barrier. Then we provide pricing model and formula for defaultable corporate bonds with discrete coupons and consider its duration and the effect of the government taxes.

The one factor pricing model between every adjacent two coupon dates becomes an inhomogeneous Black-Scholes equation with constant coefficients and discontinuous terminal value condition and can be solved by the method of higher order binaries with constant coefficients which is used to the pricing problem of corporate zero coupon bonds in [11]. The two factor pricing model between every adjacent two coupon dates can be changed to an inhomogeneous Black-Scholes equation with time dependent coefficients and discontinuous terminal value condition through change of numeraire, which can be solved by the method of higher order binaries which is used to the pricing problem of corporate zero coupon bonds on [10].

The remainder of the article is divided into two parts.

The first part describes *one factor model* and is organized as follows. In the section 2 we consider the model and pricing formula for equity value and using it calculate expected default barrier. Then we provide pricing model and formula for defaultable discrete coupon corporate bonds without consideration of taxes. In the section 3 we study duration of our bond. In the section 4 we consider the effect of taxes. The section 5 is an appendix where we give the sketch of the proof of pricing formulae for equity value and defaultable discrete coupon bond. The notions and the pricing formulae of higher order binaries with constant coefficients and their some properties which are used in the sections 2, 3 and 4 will be referred to [2], [8] or [11].

The second part describes *two factors model* and is article is organized as follows. The section 1 gives an introduction. In the section 2 we consider the model and pricing formula for equity value and using it calculate expected default barrier. Then we provide pricing model and formula for defaultable discrete coupon corporate bonds without consideration of taxes. In the section 3 we study duration of our bond. In the section 4 we consider the effect of taxes. The section 5 is an appendix where we give the sketch of the proof of pricing formulae for equity value. The notions and the pricing formulae of higher order binaries with time dependent coefficients and their some properties which are used in the sections 2, 3 and 4 will be referred to [10].

# PART I: One Factor Model

## 2. Mathematical Model and Pricing Formulae for Discrete Coupon Bond with both Expected and Unexpected Defaults

**2.1 Assumptions**

1) Short rate *r* is constant. Then the price of default free zero coupon bond with maturity *T* and face value 1 is $Z(t;T) = e^{-r(T-t)}$.

2) The firm value $V(t)$ follows a geometric Brown motion
$$dV(t) = (r-b)V(t)dt + s_V(t)V(t)dW(t)$$
under the risk neutral martingale measure. The firm continuously pays out dividend in rate $b \geq 0$ (constant) for a unit of firm value.

3) Let $0 = T_0 < T_1 < \cdots < T_{N-1} < T_N = T$ and *T* is the maturity of our corporate bond (debt) with face value *F* (unit of currency). At time $T_i$ ($i = 1,\ldots, N-1$) bond holder receives the coupon of quantity $C_i$ (unit of currency) from the firm and at time $T_N = T$ bond holder receives the face value *F* and the last coupon $C_N$ (unit of currency). (That is, the coupons are the same as in [1])

4) The expected default occurs only at time $T_i$ when the equity of the firm is not enough to pay debt and coupon. If the expected default occurs, the bond holder receives $\delta \cdot V$ as *default recovery*. Here $\delta$ is called a *fractional recovery rate* of firm value at default.

5) The unexpected default can occur at any time. The unexpected default probability in the time interval $[t, t+\Delta t] \cap [T_i, T_{i+1}]$ is $\lambda_i \Delta t$ ($i = 0, \cdots, N-1$). Here the *default intensity* $\lambda_i$ is a constant. If the unexpected default occurs at time $t \in (T_i, T_{i+1})$, the bond holder receives $\min\{\delta \cdot V, \; \Sigma_{k=i+1}^{N} C_k Z(t;T_k) + FZ(t;T_N)\}$ as default recovery. Here the *reason* why the expected default recovery and unexpected recovery are *given* in *different* forms is to avoid the possibility of *paying more than the price of default free discrete coupon bond* with the face value *F* and coupons $C_k$ (at time $T_k$) as a default recovery when the unexpected default event occurs. In what follows we call the unexpected default occurred at time $t \in (T_i, T_{i+1})$ with default recovery $\Sigma_{k=i+1}^{N} C_k Z(t;T_k) + FZ(t;T_N)$ as the *unexpected default without loss*.

6) In the subinterval $(T_i, T_{i+1}]$, the price of our corporate bond and the equity of the firm are given by a sufficiently smooth function $B_i(V,t)$ and $E_i(V,t)$ ($i = 0, \cdots, N-1$), respectively.

**2.2 Mathematical Model for Equity and Expected Default Barriers**

According to the result the subsection 2.2 of the part II, we can derive *PDE* of the *equity E* when the firm has constant default intensity $\lambda$ under the assumptions 1) and 2).

$$\frac{\partial E}{\partial t} + \frac{1}{2}s_V^2 V^2 \frac{\partial^2 E}{\partial V^2} + (r-b)V\frac{\partial E}{\partial V} - (r+\lambda)E = 0.$$

Thus we derive the mathematical model for the equity under the assumptions 1) ~ 6). From the above PDE of the equity and the above assumption 5), 6) the equity price $E_i$ satisfies the following PDE in every subinterval $(T_i, T_{i+1})$ ($i = 0, \cdots, N-1$):

$$\frac{\partial E_i}{\partial t} + \frac{1}{2} s_V^2 V^2 \frac{\partial^2 E_i}{\partial V^2} + (r-b)V \frac{\partial E_i}{\partial V} - (r + \lambda_i) E_i = 0. \tag{2.1}$$

From the assumption 3) we have:

$$\begin{aligned} E_{N-1}(V, T_N) &= (V - F - C_N) \cdot 1\{V \geq F + C_N\}, \\ E_i(V, T_{i+1}) &= [E_{i+1}(V, T_{i+1}) - C_{i+1}] \cdot 1\{E_{i+1}(V, T_{i+1}) \geq C_{i+1}\}, \quad i = 0, \cdots, N-2. \end{aligned} \tag{2.2}$$

We will use the following notation for simplicity.

$$\begin{aligned} K_N &= F + C_N; \; \bar{c}_N = F + C_N; \; \bar{c}_i = C_i, i = 1, \cdots, N-1; \\ \Delta T_i &= T_{i+1} - T_i, i = 0, \cdots, N-1. \end{aligned} \tag{2.3}$$

**Remark 1.** $\bar{c}_i$ is the time $T_N$-value of the payoff to bondholders at time $T_i$ ($i = 1, \ldots, N$) and $K_N$ denotes the default barrier at time $T_N$ as in [11].

**Theorem 1.** (Equity Price) *The solutions of* (2.1) *and* (2.2) *are provided as follows:*

$$E_i(V, t) = e^{-\lambda_i (T_{i+1} - t)} \left\{ e^{-\sum_{k=i+1}^{N-1} \lambda_k \Delta T_k} A_{K_{i+1} \cdots K_N}^{+ \cdots + +}(V, t; T_{i+1}, \cdots, T_N; r, b, s_V) \right. \tag{2.4}$$

$$\left. - \sum_{m=i}^{N-1} \bar{c}_{m+1} e^{-\sum_{k=i+1}^{m} \lambda_k \Delta T_k} B_{K_{i+1} \cdots K_{m+1}}^{+ \cdots +}(V, t; T_{i+1}, \cdots, T_{m+1}; r, b, s_V) \right\}.$$

$$(T_i < t \leq T_{i+1}, \; V > 0, \; i = 0, 1, \cdots, N-1.)$$

*Here* $B_{K_1 \cdots K_m}^{+ \cdots +}(x, t; T_1, \cdots, T_m; r, q, \sigma)$ *and* $A_{K_1 \cdots K_{m-1} K_m}^{+ \cdots + +}(x, t; T_1, \cdots, T_{m-1}, T_m; r, b, \sigma)$ *are the prices of m-th order bond and asset binaries with risk free rate **r**, dividend rate **q** and volatility $\sigma$ (the lemma 1 of [8]) and $K_i (i = 1, \cdots, N-1)$ is the unique root of the equation $E_i(V, T_i) = C_i$. Using multi-variate normal distribution functions, (2.4) are represented in terms of the debt F, the coupons $C_i$ and the firm value V as follows:*

$$\begin{aligned} E_i(V, t) &= V \cdot e^{-(\lambda_i + b)(T_{i+1} - t) - \sum_{k=i+1}^{N-1} (\lambda_k + b) \Delta T_k} N_{N-i}(d_{i+1}^+(t), \cdots, d_N^+(t); A_{i+1,N}(t)) - \\ &\quad - (F + C_N) Z(t, T_N) e^{-\lambda_i (T_{i+1} - t) - \sum_{k=i+1}^{N-1} \lambda_k \Delta T_k} N_{N-i}(d_{i+1}^-(t), \cdots, d_N^-(t); A_{i+1,N}(t)) \\ &\quad - \sum_{m=i}^{N-2} C_{m+1} Z(t, T_{m+1}) e^{-\lambda_i (T_{i+1} - t) - \sum_{k=i+1}^{m} \lambda_k \Delta T_k} N_{m+1-i}(d_{i+1}^-(t), \cdots, d_{m+1}^-(t); A_{i+1,m+1}(t)). \end{aligned} \tag{2.5}$$

*Here $N_m(a_1, \cdots, a_m; A)$ ( the cumulative distribution function of m-variate normal distribution with zero mean vector and a covariance matrix $A^{-1}$), $d_i^{\pm}(t)$ and $(A_{k,m}(t))^{-1} = (r_{ij}(t))_{i,j=k}^{m}$ are given by:*

$$N_m(a_1,\cdots,a_m\ ;A)=\int_{-\infty}^{a_1}\cdots\int_{-\infty}^{a_m}\frac{1}{(\sqrt{2\pi})^m}\sqrt{\det A}\exp(-\frac{1}{2}y^\perp Ay)dy,$$

$$d_i^\pm(t)=\left(s_V\sqrt{T_i-t}\right)^{-1/2}\left[\ln\frac{V}{K_i}+\left(r-b\pm\frac{1}{2}s_V^2\right)(T_i-t)\right],i=1,\cdots,N-1,$$

$$d_N^\pm(t)=\left(s_V\sqrt{T_N-t}\right)^{-1/2}\left[\ln\frac{V}{F+C_N}+\left(r-b\pm\frac{1}{2}s_V^2\right)(T_N-t)\right],$$

$$r_{ij}(t)=\sqrt{(T_i-t)/(T_j-t)}\ ,\ r_{ji}(t)=r_{ij}(t),\ i\le j\ (i,j=k,\cdots,m). \tag{2.6}$$

**Remark 2**. The theorem 1 gives us the expected default barrier $K_i$ at time $T_i$ ($i$ =1,…, $N$−1). That is, if $V < K_i$ at time $T_i$, then the expected default occurs. Note that the difference of (2.4) and (2.5) from (2.13), (2.14) and (2.15) of [10] comes from the coupon structures' difference. If $b$ = 0 and $\lambda_k$ = 0 ($i$ = 0,…, $N$−1), then our pricing formula (2.5) has the same type with the formula (2) of [1, at page 751] but we should *note* that here short rate *r* is constant. If $b$ = 0, $C_k$ = 0 and $\lambda_k$ = 0 ($i$ = 0,…, $N$−1), then the formula (2.5) with (2.6) includes the formula (12) of Merton (1974) [7].

### 2.3 Model and Pricing Formulae of the Defaultable Discrete Coupon Bond

In this subsection we derive the representation of the price $B_i(V, t)$ of the defaultable discrete coupon bond in the interval $(T_i, T_{i+1}]$ ($i$ = 0,…, $N$−1). In this subsection we *neglect* the effect of the *taxation*. We use the notation of (2.3) and the following notation

$$\Phi_i(t)=\Sigma_{k=i+1}^{N}C_k Z(t;T_k)+FZ(t;T_N). \tag{2.7}$$

That is, $\Phi_i(t)$ is time *t*-value of *default free discrete coupon bond* with the *maturity* $T_N$ - face value $F$ and coupons $C_{i+1},\cdots,C_N$ at time $T_{i+1},\cdots,T_N$.

Now we consider the defaultable discrete coupon bond under the assumptions 1) ~ 6) in the subsection 2.1. From the assumption 5) and 6), using the method of [13] we can know that our bond price $B_i$ satisfies the following PDE in every subinterval $(T_i, T_{i+1})$ ($i=1,\cdots,N$):

$$\frac{\partial B_i}{\partial t}+\frac{1}{2}s_V^2 V^2\frac{\partial^2 B_i}{\partial V^2}+(r-b)V\frac{\partial B_i}{\partial V}-(r+\lambda_i)B_i+\lambda_i\min\{\delta V,\Phi_i(t)\}=0, T_i<t<T_{i+1}, V>0. \tag{2.8}$$

In the theorem 1, we have calculated the expected default barrier $K_i$ ($i$ =1,…, $N$) (see the remark 2). Thus from the assumptions 3) and 4) we have the following terminal value conditions:

$$\begin{aligned}&B_{N-1}(V,T_N)=\bar{c}_N\cdot 1\{V\ge K_N\}+\delta V\cdot 1\{V<K_N\},\ V>0;\\&B_i(V,T_{i+1})=[B_{i+1}(V,T_{i+1})+\bar{c}_{i+1}]\cdot 1\{V\ge K_{i+1}\}+\delta V\cdot 1\{V<K_{i+1}\}, V>0, i=0,\cdots,N-2.\end{aligned} \tag{2.9}$$

The problem (2.8) and (2.9) is just the pricing model of our defaultable discrete coupon bond.

**Remark 3**. In our model (2.8) and (2.9) the consideration of *unexpected default risk* and dividend of firm value is added to the model on defaultable discrete coupon bond of [1]. Another difference from [1]'s approach is that risk free rate *r* is constant (but not stochastic process) . The difference from the model (2.18), (2.19) of [10] is the difference of coupon structures.

Our model (2.8) and (2.9) has some difference in default barriers and default recovery from the model (3.5) of [11] for defaultable zero coupon bond with discrete default information and endogenous default recovery but it is very similar with the fundamental problem (4.2) of [11] which is a terminal value problem for an inhomogenous Black - Scholes equation with constant coefficients and binary type terminal value.

**Theorem 2.** (Discrete Coupon Bond Price) *The solution to* (2.8) *and* (2.9) *is given by:*

$$B_i(V,t) = e^{-\lambda_i(T_{i+1}-t)} \left\{ \sum_{m=i}^{N-1} e^{-\sum_{k=i+1}^{m} \lambda_k \Delta T_k} \left[ \bar{c}_{m+1} B_{K_{i+1}\cdots K_{m+1}}^{+\cdots+}(V,t;T_{i+1},\cdots,T_{m+1};r,b,s_V) \right.\right.$$
$$\left.\left. + \delta \cdot A_{K_{i+1}\cdots K_m K_{m+1}}^{+\cdots+-}(V,t;T_{i+1},\cdots,T_m,T_{m+1};r,b,s_V) \right] \right.$$
$$+ \sum_{m=i+1}^{N-1} \lambda_m e^{-\sum_{k=i+1}^{m-1} \lambda_k \Delta T_k} \int_{T_m}^{T_{m+1}} e^{-\lambda_m(\tau-T_m)} \left[ \Phi_m(\tau) \cdot B_{K_{i+1}\cdots K_m M_{m+1}(\tau)}^{+\cdots++}(V,t;T_{i+1},\cdots,T_m,\tau;r,b,s_V) \right.$$
$$\left.\left. + \delta \cdot A_{K_{i+1}\cdots K_m M_{m+1}(\tau)}^{+\cdots+-}(V,t;T_{i+1},\cdots,T_m,\tau;r,b,s_V) \right] d\tau \right\} +$$
$$+ \lambda_i \int_t^{T_{i+1}} e^{-\lambda_i(\tau-t)} \left[ \Phi_i(\tau) B_{M_{i+1}(\tau)}^{+}(V,t;\tau;r,b,s_V) + \delta \cdot A_{M_{i+1}(\tau)}^{-}(V,t;\tau;r,b,s_V) \right] d\tau,$$

$$T_i < t \leq T_{i+1},\ V > 0,\ i = 0,\cdots,N-1. \quad (2.10)$$

*Here* $B_{K_1\cdots K_m}^{+\cdots+}$, $A_{K_1\cdots K_{m-1}K_m}^{+\cdots+-}$, $\bar{c}_i, K_i (i=1,\cdots,N)$ *are the same with the theorem 1 and* $M_{i+1}(t) = \delta^{-1}\Phi_i(t)$. *In particular the initial price of the bond is given by*

$$B_0 = B_0(V_0, 0) = \sum_{m=0}^{N-1} e^{-\sum_{k=0}^{m-1} \lambda_k \Delta T_k} \left\{ e^{-\lambda_m \Delta T_m} \left[ \bar{c}_{m+1} B_{K_1\cdots K_{m+1}}^{+\cdots+}(V_0,0;T_1,\cdots,T_{m+1};r,b,s_V) \right.\right.$$
$$\left. + \delta \cdot A_{K_1\cdots K_m K_{m+1}}^{+\cdots+-}(V_0,0;T_1,\cdots,T_m,T_{m+1};r,b,s_V) \right]$$
$$+ \lambda_m \int_{T_m}^{T_{m+1}} e^{-\lambda_m(\tau-T_m)} \left[ \Phi_m(\tau) \cdot B_{K_1\cdots K_m M_{m+1}(\tau)}^{+\cdots++}(V_0,0;T_1,\cdots,T_m,\tau;r,b,s_V) \right.$$
$$\left.\left. + \delta \cdot A_{K_1\cdots K_m M_{m+1}(\tau)}^{+\cdots+-}(V_0,0;T_1,\cdots,T_m,\tau;r,b,s_V) \right] d\tau \right\}, 0 \leq t \leq T_1,\ V > 0. \quad (2.11)$$

**Remark 4**. 1) The proof of theorem 2 is similar with the solving of (4.2) of [11] and we give a sketch of the proof in the appendix. 2) The problem (2.8) and (2.9) is an inhomogenous Black-Scholes equation with discontinuous terminal value. Thus using the results of [9], we can investigate such properties of $B_i(V,t)$ as monotonicity, boundedness or gradient estimate and so on.

Let denote the ***leverage ratio*** by $L=F/V_0$ and the *k*-th ***coupon rate*** by $c_k = C_k/F$ ($k=1,\ldots,N$). Then we have the following *representation* of the *initial price* of the our *defaultable discrete coupon bond* in terms of *leverage ratio, face value, coupon rates, default recovery rate* and *initial price of the default free zero coupon bonds with maturity* $T_k$ (coupon dates).

**Corollary 1.** *Under the assumption of theorem 2, the initial price of the bond can be represented as follows*:

$B_0 = B_0(L, F, c_1,\cdots,c_N; \delta, \lambda_0,\cdots,\lambda_{N-1}; r, b) =$

$$= F\left\{ e^{-\sum_{k=0}^{N-1}\lambda_k \Delta T_k} Z(0, T_N) N_N(d_1^-, \cdots, d_N^-; A_N) + \right.$$

$$+ \sum_{m=0}^{N-1} e^{-\sum_{k=0}^{m-1}\lambda_k \Delta T_k} \left[ e^{-\lambda_m \Delta T_m} c_{m+1} Z(0, T_{m+1}) N_{m+1}(d_1^-, \cdots, d_{m+1}^-; A_{m+1}) + \right.$$

$$\left. + \lambda_m \phi_m(0) \int_{T_m}^{T_{m+1}} e^{-\lambda_m(\tau - T_m)} N_{m+1}(d_1^-, \cdots, d_m^-, \tilde{d}_{m+1}^-(\tau, \delta); \tilde{A}_{m+1}(\tau)) d\tau \right] \quad (2.12)$$

$$+ \frac{\delta}{L} \sum_{m=0}^{N-1} e^{-\sum_{k=0}^{m-1}(\lambda_k + b)\Delta T_k} \left[ e^{-(\lambda_m + b)\Delta T_m} N_{m+1}(d_1^+, \cdots, d_m^+, -d_{m+1}^+; A_{m+1}^-) \right.$$

$$\left. \left. + \lambda_m \int_{T_m}^{T_{m+1}} e^{-(\lambda_m + b)(\tau - T_m)} N_{m+1}(d_1^+, \cdots, d_m^+, -\tilde{d}_{m+1}^+(\tau, \delta); \tilde{A}_{m+1}^-(\tau)) d\tau \right] \right\}.$$

*Here*

$$\phi_m(0) = \Phi_m(0)/F = (1 + c_N) Z(0; T_N) + \Sigma_{k=m+1}^{N-1} c_k Z(0; T_k),$$

$N_m(a_1, \cdots, a_m; A)$, $d_i^\pm = d_i^\pm(0)$ *and* $(A_m)^{-1} = [A_{1,m}(0)]^{-1} = (r_{ij})_{i,j=1}^m$ *are provided by* (2.6).

$$\tilde{d}_i^\pm(\tau, \delta) = (s_V \sqrt{\tau})^{-1/2} \left[ \ln \frac{V}{M_i(\tau)} + \left( r - b \pm \frac{1}{2} s_V^2 \right) \tau \right], T_{i-1} \leq \tau < T_i; i = 1, \cdots, N, \quad (2.13)$$

$(\tilde{A}_m(\tau))^{-1} = (\tilde{r}_{ij}(\tau))_{i,j=1}^m$ *is the matrix whose m-th row and column are given by*

$$\tilde{r}_{im}(\tau) = \sqrt{T_i/\tau}, \ \tilde{r}_{mi}(\tau) = \tilde{r}_{im}(\tau), \ i < m \ (i = 1, \cdots, m-1) \quad (2.14)$$

*and other elements coincide with those of* $(A_m)^{-1}$. *The matrices* $(A_m^-)^{-1} = (r_{ij}^-)_{i,j=1}^m$ *and* $(\tilde{A}_m^-(\tau))^{-1}$
$= (\tilde{r}_{ij}^-(\tau))_{i,j=1}^m$ *have such m-th rows and columns that*

$$r_{im}^- = -r_{im}, \ r_{mi}^- = -r_{mi}; \ \tilde{r}_{im}^-(\tau) = -\tilde{r}_{im}(\tau), \ \tilde{r}_{mi}^-(\tau) = -\tilde{r}_{mi}(\tau), \ i < m \ (i = 1, \cdots, m-1) \quad (2.15)$$

*and other elements coincide with those of* $(A_m)^{-1}$ *and* $(\tilde{A}_m(\tau))^{-1}$, *respectively* .

**Remark 5.** If $b = 0$ and $\lambda_k = 0$ ($i = 0, \ldots, N-1$), then our pricing formula (2.12) nearly coincides with the formula (5) of [1, at page 752] and the only difference comes from the assumption of short rate. If $b = 0$, $C_k = 0$ and $\lambda_k = 0$ ($i = 0, \ldots, N-1$), then the formula (2.12) includes the formula (13) of Merton (1974) [7].

In what follows, we use the following notation for simplicity:

$$G_N^+ = G_N^+(\lambda_0, \cdots, \lambda_{N-1}; b) = e^{-\sum_{k=0}^{N-1}(\lambda_k + b)\Delta T_k} N_N(d_1^+, \cdots, d_N^+; A_N),$$

$$G_{m+1}^- = G_{m+1}^-(\lambda_0, \cdots, \lambda_m) = e^{-\sum_{k=0}^{m}\lambda_k \Delta T_k} N_{m+1}(d_1^-, \cdots, d_{m+1}^-; A_{m+1})$$

$$g_{m+1}^-(\tau) = g_{m+1}^-(\tau; \delta, \lambda_0, \cdots, \lambda_m) = e^{-\lambda_m(\tau - T_m) - \sum_{k=0}^{m-1}\lambda_k \Delta T_k} N_{m+1}(d_1^-, \cdots, d_m^-, \tilde{d}_{m+1}^-(\tau, \delta); \tilde{A}_{m+1}(\tau)), \quad (2.16)$$

$$\tilde{G}_{m+1} = \tilde{G}_{m+1}(\lambda_0, \cdots, \lambda_m; b) = e^{-\sum_{k=0}^{m}(\lambda_k + b)\Delta T_k} N_{m+1}(d_1^+, \cdots, d_m^+, -d_{m+1}^+; A_{m+1}^-),$$

$$\tilde{g}_{m+1}(\tau) = \tilde{g}_{m+1}(\tau; \delta, \lambda_0, \cdots, \lambda_m; b) =$$
$$= e^{-(\lambda_m + b)(\tau - T_m) - \sum_{k=0}^{m-1}(\lambda_k + b)\Delta T_k} N_{m+1}(d_1^+, \cdots, d_m^+, -\tilde{d}_{m+1}^+(\tau, \delta); \tilde{A}_{m+1}^-(\tau)), \ m = 0, \cdots, N-1.$$

Then from (2.5) and (2.12) we can write as follows:

$$E_0(V_0, r, 0) = V_0 G_N^+ - (F + C_N)Z(0, T_N)G_N^- - \sum_{m=1}^{N-1} C_m Z(0, T_m) G_m^-,$$

$$B_0(V_0, r, 0) = (F + C_N)Z(0, T_N)G_N^- + \sum_{m=1}^{N-1} C_m Z(0, T_m) G_m^- + \sum_{m=0}^{N-1} \lambda_m \Phi_m(0) \int_{T_m}^{T_{m+1}} g_{m+1}^-(\tau) d\tau +$$

$$+ \delta V_0 \sum_{m=0}^{N-1} \left( \tilde{G}_{m+1} + \lambda_m \int_{T_m}^{T_{m+1}} \tilde{g}_{m+1}(\tau) d\tau \right). \quad (2.17)$$

If we take the sum of the above expressions, we have

$$E_0 + B_0 = V_0 G_N^+ + \delta V_0 \sum_{m=0}^{N-1} \left( \tilde{G}_{m+1} + \lambda_m \int_{T_m}^{T_{m+1}} \tilde{g}_{m+1}(\tau) d\tau \right) + \sum_{m=0}^{N-1} \lambda_m \Phi_m(0) \int_{T_m}^{T_{m+1}} g_{m+1}^-(\tau) d\tau.$$

Therefore we have

$$V_0 = E_0 + B_0 + V_0 \left[ 1 - G_N^+ - \delta \sum_{m=0}^{N-1} \left( \tilde{G}_{m+1} + \lambda_m \int_{T_m}^{T_{m+1}} \tilde{g}_{m+1}(\tau) d\tau \right) \right] - \sum_{m=0}^{N-1} \lambda_m \Phi_m(0) \int_{T_m}^{T_{m+1}} g_{m+1}^-(\tau) d\tau.$$

This shows that the Modigliani-Miller theorem holds (that is, $V = Equity + Debt$) when $\delta = 1$ and $\lambda_k = b = 0$. (Here we considered the following fact [1]:

$$1 - N_N(d_1^+, \cdots, d_N^+; A_N) = \sum_{m=0}^{N-1} N_{m+1}(d_1^+, \cdots, d_m^+, -d_{m+1}^+; A_{m+1}^-).)$$

In the case with possibility of default, it is modified as follows [1]:

$$V = Equity + Debt + Default\ Costs\ (bankruptcy\ costs).$$

From this fact, we have the representation of *bankruptcy costs*.

**Corollary 2.** (Bankruptcy Cost) *The current value of bankruptcy cost is as follows:*

$$V_0 - V_0 G_N^+ - \sum_{m=0}^{N-1} \left\{ V_0 \delta \tilde{G}_{m+1} + \lambda_m \int_{T_m}^{T_{m+1}} [V_0 \delta \tilde{g}_{m+1}(\tau) + \Phi_m(0) g_{m+1}^-(\tau)] d\tau \right\}. \quad (2.18)$$

**Remark 6.** In the formula (2.18), let $\lambda_k = b = 0$, then we have the formula (6) of [1, at page 752].

## 3. Duration

In this section we study the problem of duration for defaultable discrete coupon bond under the united model of structural and reduced form approaches we developed in the previous section. According to [1], when $B(V, t; r)$ is bond price, we use the following definition for *duration* with respect to the short rate

$$D(V, t) = -\frac{1}{B(V, t)} \partial_r B(V, t; r). \quad (3.1)$$

For example, the *duration* $D_Z(t, T)$ of *default free zero coupon bond* $Z(t; T)$ under the assumption 1) in the section 2 is just the $T - t$. The *duration* of the *default free discrete coupon bond* for $t \in [0, T_1]$, the holder of which receives the coupon of quantity $C_i$ (unit of currency) at time $T_i$ ($i = 1, \ldots, N-1$) and receives the face value $F$ and the last coupon $C_N$ (unit of currency) at time $T_N=T$, is given by

$$D(t;F,C_1,\cdots,C_N)=\frac{-\partial_r \Phi_0(t)}{\Phi_0(t)}=\frac{(F+C_N)Z(t;T_N)}{\Phi_0(t)}(T_N-t)+\sum_{k=1}^{N-1}\frac{C_k Z(t;T_k)}{\Phi_0(t)}(T_k-t), \quad (3.2)$$

because its time $t$-price is just the same with

$$\Phi_0(t)=\Sigma_{k=1}^{N-1}C_k Z(t;T_k)+(F+C_N)Z(t;T_N),\ \ t\in[0,T_1].$$

*Note* that the duration of such default free discrete coupon bond with coupon dates $T_k$ is a convex linear combination of the durations of zero coupon bonds with maturity $T_k$.

Now let calculate the duration of our defaultable discrete coupon bond. We use the notation of (2.3). In (2.17) the third term can be rewritten as follows:

$$\sum_{m=0}^{N-1}\lambda_m \Phi_m(0)\int_{T_m}^{T_{m+1}} g_{m+1}^-(\tau)d\tau = \sum_{m=0}^{N-1}\lambda_m \sum_{n=m+1}^{N}\bar{c}_n Z(0;T_n)\int_{T_m}^{T_{m+1}} g_{m+1}^-(\tau)d\tau$$

$$= \sum_{n=1}^{N}\bar{c}_n Z(0;T_n)\sum_{m=0}^{n-1}\lambda_m \int_{T_m}^{T_{m+1}} g_{m+1}^-(\tau)d\tau.$$

Then we have another more intuitional initial price representation:

$$B_0(V_0,r,0)=\sum_{n=1}^{N}\bar{c}_n Z(0;T_n)\left[G_n^- + \sum_{m=0}^{n-1}\lambda_m \int_{T_m}^{T_{m+1}} g_{m+1}^-(\tau)d\tau\right] + \delta V_0 \sum_{m=0}^{N-1}\left[\tilde{G}_{m+1}+\lambda_m \int_{T_m}^{T_{m+1}} \tilde{g}_{m+1}(\tau)d\tau\right]. \quad (3.3)$$

Here we let

$$f_n(r)=G_n^- + \sum_{m=0}^{n-1}\lambda_m \int_{T_m}^{T_{m+1}} g_{m+1}^-(\tau)d\tau,\ n=1,\cdots,N;\ h(r)=\sum_{m=0}^{N-1}\left(\tilde{G}_{m+1}+\lambda_m \int_{T_m}^{T_{m+1}} \tilde{g}_{m+1}(\tau)d\tau\right). \quad (3.4)$$

**Remark 7.** $f_n$ may be considered as the probability of no default (or unexpected default without loss) prior or at $T_n$ and $h(r)$ the probability of expected default or unexpected default with recovery $\delta V_0$.

Then $f_n$, $h>0$ and the initial price of our bond is written as follows:

$$B_0=B_0(V_0,0;r)=\sum_{n=1}^{N}\bar{c}_n Z(0;T_n)f_n(r)+\delta V_0 h(r). \quad (3.5)$$

Thus we have

$$-\partial_r B_0=\sum_{n=1}^{N}\bar{c}_n Z(0;T_n)[T_n f_n - \partial_r f_n]-\delta V_0 \partial_r h. \quad (3.6)$$

We use the lemma on derivatives of multi-variate normal distribution functions (the lemma 1 in the section 5 in the following part II) and

$$\frac{\partial}{\partial r}d_i^\pm(0)=\left(s_V \sqrt{T_i}\right)^{-1}T_i\ ;\ \frac{\partial}{\partial r}\tilde{d}_i^\pm(\tau,\delta)=\left(s_V \sqrt{\tau}\right)^{-1}\tau, i=1,\cdots,N\ ([6])$$

to get

$$\partial_r N_{m+1}(d_1^-,\cdots,d_{m+1}^-;A_{m+1})=\sum_{i=1}^{m+1}\breve{N}_{m+1,i}(d_1^-,\cdots,d_{m+1}^-;A_{m+1})\cdot\left(s_V \sqrt{T_i}\right)^{-1}T_i \geq 0,$$

$$\partial_r N_{m+1}(d_1^+,\cdots,d_m^+,-d_{m+1}^+;A_{m+1}^-)=\sum_{i=1}^{m}\breve{N}_{m+1,i}(d_1^+,\cdots,d_m^+,-d_{m+1}^+;A_{m+1}^-)\cdot\left(s_V \sqrt{T_i}\right)^{-1}T_i -$$

$$-\breve{N}_{m+1,m+1}(d_1^+,\cdots,d_m^+,-d_{m+1}^+;A_{m+1}^-)\cdot\left(s_V\sqrt{T_{m+1}}\right)^{-1}T_{m+1},$$

$$\partial_r N_{m+1}(d_1^-,\cdots,d_m^-,\tilde{d}_{m+1}^-(\tau,\delta);\tilde{A}_{m+1}(\tau)) = \sum_{i=1}^{m}\breve{N}_{m+1,i}(d_1^-,\cdots,d_m^-,\tilde{d}_{m+1}^-(\tau,\delta);\tilde{A}_{m+1}(\tau))\cdot\left(s_V\sqrt{T_i}\right)^{-1}T_i$$

$$+\breve{N}_{m+1,m+1}(d_1^-,\cdots,d_m^-,\tilde{d}_{m+1}^-(\tau,\delta);\tilde{A}_{m+1}(\tau))\cdot\left(s_V\sqrt{\tau}\right)^{-1}\tau \geq 0,$$

$$\partial_r N_{m+1}(d_1^+,\cdots,d_m^+,-\tilde{d}_{m+1}^+(\tau,\delta);\tilde{A}_{m+1}^-(\tau)) = \sum_{i=1}^{m}\breve{N}_{m+1,i}(d_1^+,\cdots,d_m^+,-\tilde{d}_{m+1}^+(\tau,\delta);\tilde{A}_{m+1}^-(\tau))\cdot\left(s_V\sqrt{T_i}\right)^{-1}T_i$$

$$-\breve{N}_{m+1,m+1}(d_1^+,\cdots,d_m^+,-\tilde{d}_{m+1}^+(\tau,\delta);\tilde{A}_{m+1}^-(\tau))\cdot\left(s_V\sqrt{\tau}\right)^{-1}\tau.$$

Here $\breve{N}_{m+1,i}$ is given in (5.14) of [10, at page 19]. From (3.14) and (2.16), we have

$$\partial_r f_n(r) = \partial_r G_n^- + \sum_{m=0}^{n-1}\lambda_m \int_{T_m}^{T_{m+1}}\partial_r g_{m+1}^-(\tau)d\tau = \sum_{i=1}^{n}T_i(D_{n,i}^- + \tilde{J}_{i-1}^-) \geq 0, \ n=1,\cdots,N; \quad (3.7)$$

$$\partial_r h(r) = \sum_{m=0}^{N-1}\left(\partial_r \tilde{G}_{m+1} + \lambda_m \int_{T_m}^{T_{m+1}}\partial_r \tilde{g}_{m+1}(\tau)d\tau\right) = \sum_{i=1}^{N}T_i(D_{N,i}^+ - \tilde{J}_{i-1}^+). \quad (3.8)$$

Here

$$D_{n,i}^- = \left(s_V\sqrt{T_i}\right)^{-1}\Bigg[e^{-\sum_{k=0}^{n-1}\lambda_k\Delta T_k}\breve{N}_{n,i}(d_1^-,\cdots,d_n^-;A_n) +$$

$$+\sum_{m=i}^{n-1}\lambda_m\int_{T_m}^{T_{m+1}}e^{-\lambda_m(\tau-T_m)-\sum_{k=0}^{m-1}\lambda_k\Delta T_k}\breve{N}_{m+1,i}(d_1^-,\cdots,d_m^-,\tilde{d}_{m+1}^-(\tau,\delta);\tilde{A}_{m+1}(\tau))d\tau\Bigg] \geq 0, i=1,\cdots,n;$$

$$\tilde{J}_m^- = \lambda_m\int_{T_m}^{T_{m+1}}e^{-\lambda_m(\tau-T_m)-\sum_{k=0}^{m-1}\lambda_k\Delta T_k}\frac{\breve{N}_{m+1,m+1}(d_1^-,\cdots,d_m^-,\tilde{d}_{m+1}^-(\tau,\delta);\tilde{A}_{m+1}(\tau))}{s_V T_{m+1}}\sqrt{\tau}d\tau \geq 0, m=0,\cdots,n-1$$

$$D_{N,i}^+ = \left(s_V\sqrt{T_i}\right)^{-1}\Bigg[\sum_{m=i}^{N-1}\Bigg(e^{-\sum_{k=0}^{m}(\lambda_k+b)\Delta T_k}\breve{N}_{m+1,i}(d_1^+,\cdots,d_m^+,-d_{m+1}^+;A_{m+1}^-) +$$

$$+\lambda_m\int_{T_m}^{T_{m+1}}e^{-(\lambda_m+b)(\tau-T_m)-\sum_{k=0}^{m-1}(\lambda_k+b)\Delta T_k}\breve{N}_{m+1,i}(d_1^+,\cdots,d_m^+,-\tilde{d}_{m+1}^+(\tau,\delta);\tilde{A}_{m+1}^-(\tau))d\tau\Bigg)$$

$$-e^{-\sum_{k=0}^{i-1}(\lambda_k+b)\Delta T_k}\breve{N}_{i,i}(d_1^+,\cdots,d_{i-1}^+,-d_i^+;A_i^-)\Bigg], \quad i=1,\cdots,N-1; \quad (3.9)$$

$$D_{N,N}^+ = -\left(s_V\sqrt{T_N}\right)^{-1}e^{-\sum_{k=0}^{N-1}(\lambda_k+b)\Delta T_k}\breve{N}_{N,N}(d_1^+,\cdots,d_{N-1}^+,-d_N^+;A_N^-) \leq 0,$$

$$\tilde{J}_m^+ = \lambda_m\int_{T_m}^{T_{m+1}}e^{-(\lambda_m+b)(\tau-T_m)-\sum_{k=0}^{m-1}(\lambda_k+b)\Delta T_k}\frac{\breve{N}_{m+1,m+1}(d_1^+,\cdots,d_m^+,-\tilde{d}_{m+1}^+(\tau,\delta);\tilde{A}_{m+1}^-(\tau))}{s_V T_{m+1}}\sqrt{\tau}d\tau \geq 0.$$

Using these notations (3.4) and (3.9), if we substitute (3.7) and (3.8) into (3.6) we have the representation of the duration of our defaultable discrete coupon bond:

$$\tilde{D} = \frac{-\partial_r B_0}{B_0} = \frac{1}{B_0}\Bigg\{\sum_{i=1}^{N}T_i\Big[\bar{c}_i Z(0;T_i)(f_i - D_{i,i}^- - \tilde{J}_{i-1}^-) - \delta V_0(D_{N,i}^+ - \tilde{J}_{i-1}^+)\Big] -$$

$$-\sum_{i=1}^{N-1}T_i\sum_{n=i+1}^{N}\bar{c}_n Z(0;T_n)(D_{n,i}^- + \tilde{J}_{i-1}^-)\Bigg\}. \quad (3.10)$$

**Remark 8.** Here it is difficult for us to say that the duration of our defaultable discrete coupon bond is always smaller or larger than the duration (3.2) of the equivalent default free bond.

## 4. Taxes on the Coupons

In this section we extend the result of the section 2 along the line of the study of [1] on the effect of government taxes that paid on the proceeds of an investment in corporate bonds.

According to [1], State income taxes are only paid on the proceeds of an investment and not on the principal. In this case the payoff to the bond holders is reduced but the equity is not changed. Thus the expected default condition is not changed and default barrier at time $T_i$ is still $K_i$ ($i =1,…,N$) as calculated in the theorem 1. It means that when the **tax rate** is $\Lambda$ ($> 0$), the *payoff to bondholders* at coupon dates is as follows:

i) At the maturity date $T_N$, $F+(1−\Lambda)C_N$ if $V_{T_N} \geq K_N$ ($=F+C_N$) (firm value is enough large to pay debt principal $F$ and coupon $C_N$); $F+(1−\Lambda)(\delta V_{T_N}−F)$ if $F/\delta \leq V_{T_N} < K_N$ (firm value is enough large to pay debt principal but not enough to pay coupon); $\delta \cdot V_{T_N}$ if $V_{T_N} < F/\delta$ (firm value is not enough large to pay even the principal, let alone the coupon). Here we should note that this structure of the payoff comes from the *implicit* assumption that $F/\delta < F+C_N$ (equally $\delta > (1+c_N)^{-1}$ or $c_N > \delta^{-1}-1$; we call it the case II) which is possible but generally unlikable because the recovery rate $\delta$ might not be able to be so large provided a coupon rate $c_N = C_N / F$ or the coupon rate $c_N$ might not be able to be so large provided a recovery rate $\delta$. For example, if $\delta = ½$, then we must have $c_N > 1$ which seems impossible. When $F/\delta \geq F + C_N$ (equally $\delta \leq (1+c_N)^{-1}$ or $c_N \leq \delta^{-1}-1$; we call it the case I) , the payoff to bondholders at the maturity date $T_N$ is $F+(1−\Lambda)C_N$ if $V_{T_N} \geq K_N$ and $\delta \cdot V_{T_N}$ if $V_{T_N} < K_N$. Here we only consider the case I as in [1].

ii) At the $k$-th coupon date $T_i$ ($i = 1,…, N−1$), $(1−\Lambda)C_i$ if $V_{T_i} \geq K_i$; $\delta \cdot V_{T_i}$ if $V_{T_i} < K_i$. (Note that it is possible to consider the case II as at time $T_N$ but we do not consider it since it is generally unlikable.)

Let modify our pricing model (2.8) and (2.9) under consideration of taxes on the coupons provided in the above. We introduce the following notation for simplicity of pricing formulae as the previous subsections.

$$\bar{c}_N = F + (1-\Lambda)C_N;\ \bar{c}_i = (1-\Lambda)C_i, \tilde{\Phi}_i(t) = \Sigma_{k=i+1}^{N}\bar{c}_k Z(t;T_k),\ i=1,\cdots,N-1. \quad (4.1)$$

That is, $\bar{c}_i$ is the time $T_N$-value of the payoff to bondholders at time $T_i$ ($i =1,…, N$) and $\tilde{\Phi}_i(t)$ is time $t$-value of *default free discrete coupon bond* with the *maturity* $T_N$ - face value $F$ and coupons $C_{i+1},\cdots,C_N$ at time $T_{i+1},\cdots,T_N$ under consideration of the tax rate $\Lambda$.

Under the above assumption and the notation (4.1), our bond price $\tilde{B}_i$ satisfies the following PDE in every subinterval $(T_i, T_{i+1})$ ($i=0,\cdots,N-1$):

$$\frac{\partial \tilde{B}_i}{\partial t} + \frac{1}{2}s_V^2 V^2 \frac{\partial^2 \tilde{B}_i}{\partial V^2} + (r-b)V\frac{\partial \tilde{B}_i}{\partial V} - (r+\lambda_i)\tilde{B}_i + \lambda_i \min\{\delta \cdot V,\ \tilde{\Phi}_i(t)\}=0, T_i < t < T_{i+1},\ V > 0. \quad (4.2)$$

If we consider the payoff to bondholders at coupon dates, we can derive the following terminal value conditions:

$$\tilde{B}_{N-1}(V, T_N) = \bar{c}_N \cdot 1\{V \geq K_N\} + \delta V \cdot 1\{V < K_N\},$$
$$\tilde{B}_i(V, T_{i+1}) = [\tilde{B}_{i+1}(V, T_{i+1}) + \bar{c}_{i+1}] \cdot 1\{V \geq K_i\} + \delta V \cdot 1\{V < K_i\}, \ V > 0, \ i = 0, \cdots, N-2.$$
(4.3)

The problem (4.2) and (4.3) with the notation (4.1) is just the *pricing model* of our *defaultable discrete coupon bond* under consideration of *taxes on coupons* and it is the same problem with (2.8) and (2.9). Thus we have the solution representation of it just as in the theorem 2.

**Theorem 3**. *Unless the coupon rates are large relative to* $1/\delta$, *under State tax rate* $\Lambda$, *we have the following representation of the initial price of the our defaultable discrete coupon bond in terms of debt, coupon rates, default recovery rate, default intensity, and initial price of the default free zero coupon bond and initial firm value:*

$$\tilde{B}_0 = \tilde{B}_0(V_0, F, c_1, \cdots, c_N; \delta, \lambda_0, \cdots, \lambda_{N-1}; r, b, \Lambda) =$$
$$= F\left\{ Z(0, T_N) e^{-\sum_{k=0}^{N-1} \lambda_k \Delta T_k} N_N(d_1^-, \cdots, d_{N-1}^-, d_N^-; A_N) + \right.$$
$$+ \sum_{m=0}^{N-1} e^{-\sum_{k=0}^{m-1} \lambda_k \Delta T_k} \left[ (1-\Lambda) c_{m+1} Z(0, T_{m+1}) e^{-\lambda_m \Delta T_m} N_{m+1}(d_1^-, \cdots, d_{m+1}^-; A_{m+1}) + \right.$$
$$\left. \left. + \lambda_m \tilde{\phi}_m(0) \int_{T_m}^{T_{m+1}} e^{-\lambda_m (\tau - T_m)} N_{m+1}(d_1^-, \cdots, d_m^-, \tilde{d}_{m+1}^-(\tau, \delta); \tilde{A}_{m+1}(\tau)) d\tau \right] \right\} +$$
$$+ \delta V_0 \sum_{m=0}^{N-1} e^{-\sum_{k=0}^{m-1} (\lambda_k + b) \Delta T_k} \left[ e^{-(\lambda_m + b) \Delta T_m} N_{m+1}(d_1^+, \cdots, d_m^+, -d_{m+1}^+; A_{m+1}^-) + \right.$$
$$\left. + \lambda_m \int_{T_m}^{T_{m+1}} e^{-(\lambda_m + b)(\tau - T_m)} N_{m+1}(d_1^+, \cdots, d_m^+, -\hat{d}_{m+1}^+(\tau, \delta, \Lambda); \tilde{A}_{m+1}^-(\tau)) d\tau \right]. \quad (4.4)$$

*Here*

$$\tilde{\phi}_m(0) = \tilde{\Phi}_m(0)/F = Z(0; T_N) + \Sigma_{k=m+1}^N (1-\Lambda) c_k Z(0; T_k),$$

$N_m(a_1, \cdots, a_m; A), A_m, A_m^-, \tilde{A}_m(\tau), \tilde{A}_m^-(\tau)$ *and* $d_i^{\pm}$ *are the same as in the theorem 2 and* $\hat{d}_i^{\pm}(\tau, \delta, \Lambda)$ *are given by*

$$\tilde{d}_i^{\pm}(\tau, \delta) = (s_V \sqrt{\tau})^{-1/2} \left[ \ln \frac{V}{\tilde{M}_i(\tau)} + \left(r - b \pm \frac{1}{2} s_V^2\right) \tau \right], T_{i-1} \leq \tau < T_i; i = 1, \cdots, N,$$
$$\tilde{M}_{i+1}(t) = \delta^{-1} \tilde{\Phi}_i(t). \quad (4.5)$$

**Remark 9.** If $b = 0$ and $\lambda_k = 0$ ($i = 0, \ldots, N-1$), then our pricing formula (4.4) nearly coincides with the formula (10) of [1, at page 756] and the only difference comes from the fact that the short rate is constant in our model.

**Remark 10.** As in the section 2, the equation of the problem (4.2) and (4.3) is an inhomogenous Black-Scholes equation with discontinuous terminal value. Thus using the results of [10], we can investigate such properties of $\tilde{B}_i(V, t)$ as monotonicity, boundedness or gradient estimate and so on.

# 5. Appendix

## 5.1. The Proof of Theorem 1

Now we solve the problem (2.1) and (2.2). Under the notation (2.3), when i=N−1, we have

$$\frac{\partial E_{N-1}}{\partial t} + \frac{1}{2} s_V^2 V^2 \frac{\partial^2 E_{N-1}}{\partial V^2} + (r-b)V \frac{\partial E_{N-1}}{\partial V} - (r + \lambda_{N-1}) E_{N-1} = 0, T_{N-1} < t < T_N, V > 0, \quad (5.1)$$

$$E_{N-1}(V, T_N) = (V - \bar{c}_N) \cdot 1\{V \geq K_N\}, V > 0. \quad (5.2)$$

The equation (5.1) is the Black-Scholes equation with the short rate $r + \lambda_{N-1}$, the dividend rate $\lambda_{N-1} + b$ and the volatility $s_V$. The terminal value condition (5.2) can be written as

$$E_{N-1}(V, T_N) = V \cdot 1\{V \geq K_N\} - \bar{c}_N \cdot 1\{V \geq K_N\}.$$

This is the terminal value of binary option ([2, 8]) and thus we have the solution – representation in terms of binary options:

$$E_{N-1}(V,t) = A^+_{K_N}(V,t;T_N; r+\lambda_{N-1}, \lambda_{N-1}+b, s_V) - \bar{c}_N B^+_{K_N}(V,t;T_N; r+\lambda_{N-1}, \lambda_{N-1}+b, s_V)$$
$$= e^{-\lambda_{N-1}(T_N - t)}[A^+_{K_N}(V,t;T_N; r, b, s_V) - \bar{c}_N B^+_{K_N}(V,t;T_N; r, b, s_V)], T_{N-1} < t \leq T_N, V > 0. \quad (5.3)$$

Here $A^+_K(x,t;T_N;r,q,\sigma)$, $B^+_K(x,t;T_N;r,q,\sigma)$ are the prices of the asset and bond binary options with the coefficients *r*, the dividend rate *q* and the volatility $\sigma$ (the lemma1 of [11, page 4]) and we used (2.11) of [11]. In particular for the next step of study we rewrite $E_{N-1}(V, T_{N-1})$ as

$$E_{N-1}(V, T_{N-1}) = e^{-(\lambda_{N-1} - \lambda_{N-2})(T_N - T_{N-1})}[A^+_{K_N}(V, T_{N-1}; T_N; r + \lambda_{N-2}, \lambda_{N-2}+b, s_V) -$$
$$- \bar{c}_N B^+_{K_N}(V, T_{N-1}; T_N; r+\lambda_{N-2}, \lambda_{N-2}+b, s_V)]. \quad (5.4)$$

By (5.2) and the lemma 1 of [8, at page 253] we have

$$0 < \frac{\partial E_{N-1}}{\partial V}(V, T_{N-1}) < e^{-(\lambda_{N-1}+b)\Delta T_{N-1}} \leq 1, V > 0. \quad (5.5)$$

Now consider the case when *i* = N−2. In this case, (2.1) and (2.2) become

$$\frac{\partial E_{N-2}}{\partial t} + \frac{1}{2} s_V^2 V^2 \frac{\partial^2 E_{N-2}}{\partial V^2} + (r-b)V \frac{\partial E_{N-2}}{\partial V} - (r+\lambda_{N-2})E_{N-2} = 0, T_{N-2} < t < T_{N-1}, V > 0, \quad (5.6)$$

$$E_{N-2}(V, T_{N-1}) = [E_{N-1}(V, T_{N-1}) - \bar{c}_{N-1}] \cdot 1\{E_{N-1}(V, T_{N-1}) \geq \bar{c}_{N-1}\}, V > 0. \quad (5.7)$$

The equation (5.6) is the Black-Scholes equation with the short rate $r + \lambda_{N-2}$, the dividend rate $\lambda_{N-2} + b$ and the volatility $s_V$. From (5.7) the equation $E_{N-1}(V, T_{N-1}) = \bar{c}_{N-1}$ has unique root $K_{N-1}$ and $E_{N-1}(V, T_{N-1}) \geq \bar{c}_{N-1} \Leftrightarrow V \geq K_{N-1}$. (Note that $\bar{c}_{N-1} = 0 \Leftrightarrow K_{N-1} = 0$.) Thus by (5.4) the terminal value condition (5.7) can be written as follows:

$$E_{N-2}(V, T_{N-1}) = E_{N-1}(V, T_{N-1}) \cdot 1\{V \geq K_{N-1}\} - \bar{c}_{N-1} \cdot 1\{V \geq K_{N-1}\} =$$
$$= e^{-(\lambda_{N-1}-\lambda_{N-2})(T_N - T_{N-1})}[A^+_{K_N}(V, T_{N-1}; T_N; r+\lambda_{N-2}, \lambda_{N-2}+b, s_V) \cdot 1\{V \geq K_{N-1}\} -$$
$$- \bar{c}_N B^+_{K_N}(V, T_{N-1}; T_N; r+\lambda_{N-2}, \lambda_{N-2}+b, s_V) \cdot 1\{V \geq K_{N-1}\}] - \bar{c}_{N-1} \cdot 1\{V \geq K_{N-1}\}.$$

This is the terminal value of combination of the second order binaries and bond binary ([2, 8]) with the short rate $r + \lambda_{N-2}$, the dividend rate $\lambda_{N-2} + b$ and the volatility $s_V$. Thus we have the following representation:

$$E_{N-2}(V,t) = e^{-(\lambda_{N-1}-\lambda_{N-2})\Delta T_{N-1}}[A^{++}_{K_{N-1} K_N}(V,t; T_{N-1}, T_N; r+\lambda_{N-2}, \lambda_{N-2}+b, s_V)$$
$$- \bar{c}_N B^{++}_{K_{N-1} K_N}(V,t; T_{N-1}, T_N; r+\lambda_{N-2}, \lambda_{N-2}+b, s_V)]$$

$$-\bar{c}_{N-1} B^+_{K_{N-1}}(V,t;T_{N-1};r+\lambda_{N-2},\lambda_{N-2}+b,s_V) =$$
$$= e^{-\lambda_{N-2}(T_{N-1}-t)-\lambda_{N-1}\Delta T_{N-1}}[A^{++}_{K_{N-1}K_N}(V,t;T_{N-1},T_N;r,b,s_V) - \bar{c}_N B^{++}_{K_{N-1}K_N}(V,t;T_{N-1},T_N;r,b,s_V)] \quad (5.8)$$
$$- e^{-\lambda_{N-2}(T_{N-1}-t)} \bar{c}_{N-1} B^+_{K_{N-1}}(V,t;T_{N-1};r,b,s_V), \quad T_{N-2} < t \le T_{N-1},\ V > 0.$$

Here $A^{++}_{KM}(x,t;T_{N-1},T_N;r,q,\sigma)$, $B^{++}_{KM}(x,t;T_{N-1},T_N;r,q,\sigma)$ are the prices of the second order asset and bond binary options with the coefficients $r$, the dividend rate $q$ and the volatility $\sigma$ (lemma1 of [11, page 4]) and we used (2.11) of [11]. In particular for the next step of study we rewrite $E_{N-2}(V, T_{N-2})$ as

$$E_{N-2}(V, T_{N-2}) =$$
$$= e^{-\lambda_{N-1}\Delta T_{N-1}-\lambda_{N-2}\Delta T_{N-2}+\lambda_{N-3}(T_N-T_{N-2})}[A^{++}_{K_{N-1}K_N}(V,T_{N-2};T_{N-1},T_N;r+\lambda_{N-3},\lambda_{N-3}+b,s_V)$$
$$- \bar{c}_N B^{++}_{K_{N-1}K_N}(V,T_{N-2};T_{N-1},T_N;r+\lambda_{N-3},\lambda_{N-3}+b,s_V)] \quad (5.9)$$
$$- e^{-\lambda_{N-2}\Delta T_{N-2}+\lambda_{N-3}\Delta T_{N-2}} \bar{c}_{N-1} B^+_{K_{N-1}}(V,T_{N-2};T_{N-1};r+\lambda_{N-3},\lambda_{N-3}+b,s_V).$$

By (5.5), (5.7) and the lemma 1 of [8, at page 253] we have

$$0 < \frac{\partial E_{N-2}}{\partial V}(V,T_{N-2}) < e^{-(\lambda_{N-2}+b)\Delta T_{N-2}-(\lambda_{N-1}+b)\Delta T_{N-1}} \le 1,\ V > 0. \quad (5.10)$$

Now consider the case when $i = N–3$. In this case, (2.1) and (2.2) become

$$\frac{\partial E_{N-3}}{\partial t} + \frac{1}{2} s_V^2 V^2 \frac{\partial^2 E_{N-3}}{\partial V^2} + (r-b)V \frac{\partial E_{N-3}}{\partial V} - (r+\lambda_{N-3})E_{N-3} = 0,\ T_{N-3} < t < T_{N-2}, V > 0, \quad (5.11)$$

$$E_{N-3}(V, T_{N-2}) = [E_{N-2}(V, T_{N-2}) - \bar{c}_{N-2}] \cdot 1\{E_{N-2}(V, T_{N-2}) \ge \bar{c}_{N-2}\},\ V > 0. \quad (5.12)$$

The equation (5.11) is the Black-Scholes equation with the short rate $r+\lambda_{N-3}$, the dividend rate $\lambda_{N-3}+b$ and the volatility $s_V$. From (5.10) the equation $E_{N-2}(V,T_{N-2}) = \bar{c}_{N-2}$ has unique root $K_{N-2}$ and $E_{N-2}(V,T_{N-2}) \ge \bar{c}_{N-2} \Leftrightarrow V \ge K_{N-2}$. (Note that $\bar{c}_{N-2} = 0 \Leftrightarrow K_{N-2} = 0$.) Thus by (5.9) the terminal value condition (5.12) can be written as follows:

$$E_{N-3}(V, T_{N-2}) = [E_{N-2}(V, T_{N-2}) - \bar{c}_{N-2}] \cdot 1\{V \ge K_{N-2}\} =$$
$$= e^{-\lambda_{N-1}\Delta T_{N-1}-\lambda_{N-2}\Delta T_{N-2}+\lambda_{N-3}(T_N-T_{N-2})}[A^{++}_{K_{N-1}K_N}(V,T_{N-2};T_{N-1},T_N;r+\lambda_{N-3},\lambda_{N-3}+b,s_V) \cdot 1\{V \ge K_{N-2}\}$$
$$- \bar{c}_N B^{++}_{K_{N-1}K_N}(V,T_{N-2};T_{N-1},T_N;r+\lambda_{N-3},\lambda_{N-3}+b,s_V) \cdot 1\{V \ge K_{N-2}\}] - \bar{c}_{N-2} 1\{V \ge K_{N-2}\}$$
$$- e^{-\lambda_{N-2}\Delta T_{N-2}+\lambda_{N-3}\Delta T_{N-2}} \bar{c}_{N-1} B^+_{K_{N-1}}(V,T_{N-2};T_{N-1};r+\lambda_{N-3},\lambda_{N-3}+b,s_V) \cdot 1\{V \ge K_{N-2}\}.$$

This is a linear combination of the terminal values of third or lower order binary options with the short rate $r+\lambda_{N-3}$, the dividend rate $\lambda_{N-3}+b$ and the volatility $s_V$ in the meaning of [8] and the solution $E_{N-3}(V,t)$ is given by the third or lower order binary options:

$$E_{N-3}(V,t) = e^{-\lambda_{N-1}\Delta T_{N-1}-\lambda_{N-2}\Delta T_{N-2}+\lambda_{N-3}(T_N-T_{N-2})}[A^{+++}_{K_{N-2}K_{N-1}K_N}(V,t;T_{N-2},T_{N-1},T_N;r+\lambda_{N-3},\lambda_{N-3}+b,s_V)$$
$$- \bar{c}_N B^{+++}_{K_{N-2}K_{N-1}K_N}(V,t;T_{N-2},T_{N-1},T_N;r+\lambda_{N-3},\lambda_{N-3}+b,s_V)]$$
$$- \bar{c}_{N-1} e^{-\lambda_{N-2}\Delta T_{N-2}+\lambda_{N-3}\Delta T_{N-2}} B^{++}_{K_{N-2}K_{N-1}}(V,t;T_{N-2},T_{N-1};r+\lambda_{N-3},\lambda_{N-3}+b,s_V)$$
$$- \bar{c}_{N-2} B^+_{K_{N-2}}(V,t;T_{N-2};r+\lambda_{N-3},\lambda_{N-3}+b,s_V)$$

$$= e^{-\lambda_{N-1}\Delta T_{N-1} - \lambda_{N-2}\Delta T_{N-2} - \lambda_{N-3}(T_{N-2}-t)}[A^+_{K_{N-2}K_{N-1}K_N}(V,t;T_{N-2},T_{N-1},T_N;r,b,s_V)$$
$$- \bar{c}_N B^+_{K_{N-2}K_{N-1}K_N}(V,t;T_{N-2},T_{N-1},T_N;r,b,s_V)]$$
$$- \bar{c}_{N-1} e^{-\lambda_{N-2}\Delta T_{N-2} - \lambda_{N-3}(T_{N-2}-t)} B^+_{K_{N-2}K_{N-1}}(V,t;T_{N-2},T_{N-1};r,b,s_V)$$
$$- \bar{c}_{N-2} e^{-\lambda_{N-3}(T_{N-2}-t)} B^+_{K_{N-2}}(V,t;T_{N-2};r,b,s_V), \quad T_{N-3} < t \leq T_{N-2}, \; V > 0.$$
(5.13)

Here $A^+_{K_{N-2}K_{N-1}K_N}, B^+_{K_{N-2}K_{N-1}K_N}$ are the prices of the third order asset and bond binary options (the theorem 1 of [8] or the lemma1 of [11, page 4]) and we used (2.11) of [11].

By induction the formulae (2.4) are proved.(QED)

### 5.2. The Proof of Theorem 2

Now we solve the problem (2.8) and (2.9). The equation (2.8) is an inhomogenous Black-Scholes equation with the short rate $r + \lambda_i$, the dividend rate $\lambda_i + b$, the volatility $s_V$ and the inhomogenous term

$$g_i(t, V) = \lambda_i \min\{\delta V, \Phi_i(t)\} = \lambda_i[\Phi_i(t) \cdot 1\{V \geq M_{i+1}(t)\} + \delta V \cdot 1\{V < M_{i+1}(t)\}], \; i = 0, \cdots, N-1. \quad (5.14)$$

Here
$$M_{i+1}(t) = \delta^{-1}\Phi_i(t) = \delta^{-1}[\Sigma_{k=i+1}^N C_k e^{-r(T_k - t)} + F e^{-r(T_N - t)}]. \quad (5.15)$$

When $i = N-1$, we have

$$\frac{\partial B_{N-1}}{\partial t} + \frac{1}{2} s_V^2 V^2 \frac{\partial^2 B_{N-1}}{\partial V^2} + (r-b)V \frac{\partial B_{N-1}}{\partial V} - (r + \lambda_{N-1})B_{N-1} + \lambda_{N-1} \min\{\delta V, \Phi_{N-1}(t)\} = 0, \quad (5.16)$$
$$T_{N-1} < t < T_N, V > 0,$$

$$B_{N-1}(V, T_N) = \bar{c}_N \cdot 1\{V \geq K_N\} + \delta V \cdot 1\{V < K_N\}, \; V > 0. \quad (5.17)$$

The solution of (5.16) and (5.17) is given by the sum of the following two problems:

$$\frac{\partial X}{\partial t} + \frac{1}{2} s_V^2 V^2 \frac{\partial^2 X}{\partial V^2} + (r-b)V \frac{\partial X}{\partial V} - (r + \lambda_{N-1})X = 0, \quad T_{N-1} < t < T_N, V > 0, \quad (5.18)$$

$$X(V, T_N) = \bar{c}_N \cdot 1\{V \geq K_N\} + \delta V \cdot 1\{V < K_N\}, \; V > 0. \quad (5.19)$$

$$\frac{\partial Y}{\partial t} + \frac{1}{2} s_V^2 V^2 \frac{\partial^2 Y}{\partial V^2} + (r-b)V \frac{\partial Y}{\partial V} - (r + \lambda_{N-1})Y + g_{N-1}(t,V) = 0, T_{N-1} < t < T_N, V > 0, \quad (5.20)$$

$$Y(V, T_N) = 0, \; V > 0. \quad (5.21)$$

The problem (5.18) and (5.19) is binary option pricing problem with the short rate $r + \lambda_{N-1}$, the dividend rate $\lambda_{N-1} + b$ and the volatility $s_V$. Thus using the notation and binary option pricing formulae of [2 or 8] we have

$$X = \bar{c}_N B^+_{K_N}(V,t;T_N;r + \lambda_{N-1}, \lambda_{N-1} + b, s_V) + \delta A^-_{K_N}(V,t;T_N;r + \lambda_{N-1}, \lambda_{N-1} + b, s_V),$$
$$T_{N-1} \leq t < T, V > 0.$$

The problem (5.20) and (5.21) is a 0-terminal value problem of an inhomogeneous equation and thus we use the *Duhamel's principle* to solve it (see [11]). Fix $\tau \in (T_{N-1}, T_N]$ and let $W(x,t;\tau)$ be the solution to the following terminal value problem:

$$\frac{\partial W}{\partial t} + \frac{1}{2}s_V^2 V^2 \frac{\partial^2 W}{\partial V^2} + (r-b)V\frac{\partial W}{\partial V} - (r+\lambda_{N-1})W = 0, T_{N-1} < t < \tau, V > 0,$$

$$W(V,\tau;\tau) = g_{N-1}(\tau,V), \ V > 0.$$

Consider (5.14) with i=N−1 and use again the notation and binary option pricing formulae of [2 or 8] we have

$$W(V,t;\tau) = \lambda_{N-1}[\Phi_{N-1}(\tau)B^+_{M_N(\tau)}(V,t;\tau;r+\lambda_{N-1},\lambda_{N-1}+b,s_V) +$$

$$+ \delta A^-_{M_N(\tau)}(V,t;\tau;r+\lambda_{N-1},\lambda_{N-1}+b,s_V), \ T_{N-1} < t \le \tau, V > 0.$$

Then the solution $Y$ to the problem (5.20) and (5.21) is given as follows:

$$Y(V,t) = \lambda_{N-1}\int_t^{T_N}\left[\Phi_{N-1}(\tau)B^+_{M_N(\tau)}(V,t;\tau;r+\lambda_{N-1},\lambda_{N-1}+b,s_V) +\right.$$

$$\left. + \delta A^-_{M_N(\tau)}(V,t;\tau;r+\lambda_{N-1},\lambda_{N-1}+b,s_V)\right]d\tau, \quad (T_{N-1} < t \le T, V > 0).$$

Therefore $B_{N-1}(V,t), T_{N-1} < t \le T_N$ is provided as follows:

$$B_{N-1}(V,t) = \bar{c}_N B^+_{K_N}(V,t;T_N;r+\lambda_{N-1},\lambda_{N-1}+b,s_V) + \delta A^-_{K_N}(V,t;T_N;r+\lambda_{N-1},\lambda_{N-1}+b,s_V) +$$

$$+ \lambda_{N-1}\int_t^{T_N}\left[\Phi_{N-1}(\tau)B^+_{M_N(\tau)}(V,t;\tau;r+\lambda_{N-1},\lambda_{N-1}+b,s_V) + \delta A^-_{M_N(\tau)}(V,t;\tau;r+\lambda_{N-1},\lambda_{N-1}+b,s_V)\right]d\tau$$

$$= e^{-\lambda_{N-1}(T_N-t)}[\bar{c}_N B^+_{K_N}(V,t;T_N;r,b,s_V) + \delta A^-_{K_N}(V,t;T_N;r,b,s_V)] +$$

$$+ \lambda_{N-1}\int_t^{T_N} e^{-\lambda_{N-1}(\tau-t)}\left[\Phi_{N-1}(\tau)B^+_{M_N(\tau)}(V,t;\tau;r,b,s_V) + \delta A^-_{M_N(\tau)}(V,t;\tau;r,b,s_V)\right]d\tau, T_{N-1} \le t < T, V > 0.$$

(5.22)

Here the last equality comes from (2.11) of [11].

Now we consider the case when i=N−2. Then the problem (2.8) and (2.9) becomes

$$\frac{\partial B_{N-2}}{\partial t} + \frac{1}{2}s_V^2 V^2 \frac{\partial^2 B_{N-2}}{\partial V^2} + (r-b)V\frac{\partial B_{N-2}}{\partial V} - (r+\lambda_{N-2})B_{N-2} + g_{N-2}(t,V) = 0,$$

$$T_{N-2} < t < T_{N-1}, V > 0. \quad (5.23)$$

$$B_{N-2}(V,T_{N-1}) = [B_{N-1}(V,T_{N-1}) + \bar{c}_{N-1}]\cdot 1\{V \ge K_{N-1}\} + \delta V \cdot 1\{V < K_{N-1}\}, V > 0. \quad (5.24)$$

The solution to (5.23) and (5.24) is provided by the sum of the following two problems:

$$\frac{\partial X}{\partial t} + \frac{1}{2}s_V^2 V^2 \frac{\partial^2 X}{\partial V^2} + (r-b)V\frac{\partial X}{\partial V} - (r+\lambda_{N-2})X = 0, \quad T_{N-2} < t < T_{N-1}, V > 0, \quad (5.25)$$

$$X(V,T_{N-1}) = [B_{N-1}(V,T_{N-1}) + \bar{c}_{N-1}]\cdot 1\{V \ge K_{N-1}\} + \delta V \cdot 1\{V < K_{N-1}\}, V > 0. \quad (5.26)$$

$$\frac{\partial Y}{\partial t} + \frac{1}{2}s_V^2 V^2 \frac{\partial^2 Y}{\partial V^2} + (r-b)V\frac{\partial Y}{\partial V} - (r+\lambda_{N-2})Y + g_{N-2}(t,V) = 0, T_{N-2} < t < T_{N-1}, V > 0, \quad (5.27)$$

$$Y(V,T_{N-1}) = 0, \ V > 0. \quad (5.28)$$

The problem (5.27) and (5.28) is the same type with the problem (5.20) and (5.21) and thus the solution to (5.27) and (5.28) is provided as follows:

$$Y(V,t) = \lambda_{N-2}\int_t^{T_{N-1}}\left[\Phi_{N-2}(\tau)B^+_{M_{N-1}(\tau)}(V,t;\tau;r+\lambda_{N-2},\lambda_{N-2}+b,s_V) +\right.$$

$$\left. + \delta A^-_{M_{N-1}(\tau)}(V,t;\tau;r+\lambda_{N-2},\lambda_{N-2}+b,s_V)\right]d\tau, \quad (T_{N-2} < t \le T_{N-1}, V > 0).$$

Since (5.25) is an inhomogeneous Black-Scholes equation with the short rate $r+\lambda_{N-2}$, the dividend rate $\lambda_{N-2}+b$ and the volatility $s_V$, we use (2.11) of [11] to rewrite $B_{N-1}(V,T_{N-1})$ as

$$B_{N-1}(V,T_{N-1}) = e^{-(\lambda_{N-1}-\lambda_{N-2})\Delta T_{N-1}}[\,\overline{c}_N B^+_{K_N}(V,T_{N-1};T_N;r+\lambda_{N-2},b+\lambda_{N-2},s_V)$$
$$+\delta A^-_{K_N}(V,T_{N-1};T_N;r+\lambda_{N-2},b+\lambda_{N-2},s_V)]+$$
$$+\lambda_{N-1}\int_{T_{N-1}}^{T_N} e^{-(\lambda_{N-1}-\lambda_{N-2})(\tau-T_{N-1})}\Big[\Phi_{N-1}(\tau)B^+_{M_N(\tau)}(V,T_{N-1};\tau;r+\lambda_{N-2},b+\lambda_{N-2},s_V)+$$
$$+\delta A^-_{M_N(\tau)}(V,T_{N-1};\tau;r+\lambda_{N-2},b+\lambda_{N-2},s_V)\Big]d\tau.$$

Thus (5.26) can be written as

$$X(V,T_{N-1}) = e^{-(\lambda_{N-1}-\lambda_{N-2})\Delta T_{N-1}}[\,\overline{c}_N B^+_{K_N}(V,T_{N-1};T_N;r+\lambda_{N-2},b+\lambda_{N-2},s_V)\cdot 1\{V\geq K_{N-1}\}$$
$$+\delta A^-_{K_N}(V,T_{N-1};T_N;r+\lambda_{N-2},b+\lambda_{N-2},s_V)\cdot 1\{V\geq K_{N-1}\}]+$$
$$+\lambda_{N-1}\int_{T_{N-1}}^{T_N} e^{-(\lambda_{N-1}-\lambda_{N-2})(\tau-T_{N-1})}\Big[\Phi_{N-1}(\tau)B^+_{M_N(\tau)}(V,T_{N-1};\tau;r+\lambda_{N-2},b+\lambda_{N-2},s_V)\cdot 1\{V\geq K_{N-1}\}+$$
$$+\delta A^-_{M_N(\tau)}(V,T_{N-1};\tau;r+\lambda_{N-2},b+\lambda_{N-2},s_V)\cdot 1\{V\geq K_{N-1}\}\Big]d\tau+\overline{c}_{N-1}\cdot 1\{V\geq K_{N-1}\}+\delta V\cdot 1\{V<K_{N-1}\}.$$

This is a linear combination of second or lower order binaries and therefore using the notation and second order binary option pricing formulae of [2 or 8] we get the solution to the problem (5.25) and (5.26) as follows:

$$X(V,t) = e^{-(\lambda_{N-1}-\lambda_{N-2})\Delta T_{N-1}}[\,\overline{c}_N B^{++}_{K_{N-1}K_N}(V,t;T_{N-1},T_N;r+\lambda_{N-2},b+\lambda_{N-2},s_V)$$
$$+\delta A^{+-}_{K_{N-1}K_N}(V,t;T_{N-1},T_N;r+\lambda_{N-2},b+\lambda_{N-2},s_V)]+$$
$$+\lambda_{N-1}\int_{T_{N-1}}^{T_N} e^{-(\lambda_{N-1}-\lambda_{N-2})(\tau-T_{N-1})}\Big[\Phi_{N-1}(\tau)B^{++}_{K_{N-1}M_N(\tau)}(V,t;T_{N-1},\tau;r+\lambda_{N-2},b+\lambda_{N-2},s_V)+$$
$$+\delta A^{+-}_{K_{N-1}M_N(\tau)}(V,t;T_{N-1},\tau;r+\lambda_{N-2},b+\lambda_{N-2},s_V)\Big]d\tau$$
$$+\overline{c}_{N-1}\cdot B^+_{K_{N-1}}(V,t;T_{N-1};r+\lambda_{N-2},b+\lambda_{N-2},s_V)+\delta A^-_{K_{N-1}}(V,t;T_{N-1};r+\lambda_{N-2},b+\lambda_{N-2},s_V),$$
$$T_{N-2}<t\leq T_{N-1},\; V>0.$$

Thus we have the representation of $B_{N-2}(V,t)$:

$$B_{N-2}(V,t) = e^{-(\lambda_{N-1}-\lambda_{N-2})\Delta T_{N-1}}[\,\overline{c}_N B^{++}_{K_{N-1}K_N}(V,t;T_{N-1},T_N;r+\lambda_{N-2},b+\lambda_{N-2},s_V)$$
$$+\delta A^{+-}_{K_{N-1}K_N}(V,t;T_{N-1},T_N;r+\lambda_{N-2},b+\lambda_{N-2},s_V)]+$$
$$+\overline{c}_{N-1}\cdot B^+_{K_{N-1}}(V,t;T_{N-1};r+\lambda_{N-2},b+\lambda_{N-2},s_V)+\delta A^-_{K_{N-1}}(V,t;T_{N-1};r+\lambda_{N-2},b+\lambda_{N-2},s_V)$$
$$+\lambda_{N-1}\int_{T_{N-1}}^{T_N} e^{-(\lambda_{N-1}-\lambda_{N-2})(\tau-T_{N-1})}\Big[\Phi_{N-1}(\tau)B^{++}_{K_{N-1}M_N(\tau)}(V,t;T_{N-1},\tau;r+\lambda_{N-2},b+\lambda_{N-2},s_V)+$$
$$+\delta A^{+-}_{K_{N-1}M_N(\tau)}(V,t;T_{N-1},\tau;r+\lambda_{N-2},b+\lambda_{N-2},s_V)\Big]d\tau$$
$$+\lambda_{N-2}\int_t^{T_{N-1}}\Big[\Phi_{N-2}(\tau)B^+_{M_{N-1}(\tau)}(V,t;\tau;r+\lambda_{N-2},\lambda_{N-2}+b,s_V)+$$
$$+\delta A^-_{M_{N-1}(\tau)}(V,t;\tau;r+\lambda_{N-2},\lambda_{N-2}+b,s_V)\Big]d\tau,\quad (T_{N-2}<t\leq T_{N-1},\;V>0)$$
$$= e^{-\lambda_{N-2}(T_{N-1}-t)}\Big\{e^{-\lambda_{N-1}\Delta T_{N-1}}[\,\overline{c}_N B^{++}_{K_{N-1}K_N}(V,t;T_{N-1},T_N;r,b,s_V)+\delta A^{+-}_{K_{N-1}K_N}(V,t;T_{N-1},T_N;r,b,s_V)]+$$
$$+\overline{c}_{N-1}\cdot[B^+_{K_{N-1}}(V,t;T_{N-1};r,b,s_V)+\delta A^-_{K_{N-1}}(V,t;T_{N-1};r,b,s_V)]$$

$$+ \lambda_{N-1} \int_{T_{N-1}}^{T_N} e^{-\lambda_{N-1}(\tau - T_{N-1})} \Big[ \Phi_{N-1}(\tau) B^{+\ +}_{K_{N-1} M_N(\tau)}(V, t; T_{N-1}, \tau; r, b, s_V) +$$
$$+ \delta A^{+\ -}_{K_{N-1} M_N(\tau)}(V, t; T_{N-1}, \tau; r, b, s_V) \Big] d\tau \Big\} \quad (5.29)$$
$$+ \lambda_{N-2} \int_{t}^{T_{N-1}} e^{-\lambda_{N-2}(\tau - t)} \Big[ \Phi_{N-2}(\tau) B^{+}_{M_{N-1}(\tau)}(V, t; \tau; r, b, s_V) + \delta A^{-}_{M_{N-1}(\tau)}(V, t; \tau; r, b, s_V) \Big] d\tau,$$
$$(T_{N-2} < t \leq T_{N-1}, V > 0).$$

By induction we have the rest of the proof. (QED)

# PART II:  Two Factors - Model

## 1. Introduction

In the part I, we considered the one factor model for defaultable discrete coupon bond under constant short rate while Agliardi [1] considered two factors structural model. In generalizing Agliardi's two factors structural model into the unified model of structural and reduced form models, we feel *difficulties* to *get analytical pricing formulae*. The main difficulty comes from the fact that in different time intervals we must use different numeraire, which prevent us to get comprehensive formulae using higher order binaries. To overcome this difficulty we assumed that the bond holders receive the *discounted value of the predetermined quantities at the maturity* at predetermined coupon dates and the face value (debt) and the coupon at the maturity. In this case the coupons prior the maturity becomes *random* variables. Under such assumption, we can comprehensive two factor pricing formulae for defaultable discrete coupon bonds

In part II, we use independent numbers of sections, theorems and equations.

## 2. Two Factors-Model and Pricing Formulae for Discrete Coupon Bond with both Expected and Unexpected Defaults

### 2.1 Assumptions

1) Short rate follows the law
$$dr_t = a_r(r, t)dt + s_r(t)dW_1(t), \qquad a_r(r, t) = a_1(t) - a_2(t)r \quad (2.1)$$
under the risk neutral martingale measure and a standard Wiener process $W_1$. Under this assumption, the price $Z(r, t; T)$ of default free zero coupon bond is the solution to the following problem
$$\begin{cases} \dfrac{\partial Z}{\partial t} + \dfrac{1}{2} s_r^2(t) \dfrac{\partial^2 Z}{\partial r^2} + a_r(r,t) \dfrac{\partial Z}{\partial r} - rZ = 0, \\ Z(r,T) = 1. \end{cases} \quad (2.2)$$

The solution is given by

$$Z(r,t\,;T) = e^{A(t,T)-B(t,T)r}. \tag{2.3}$$

Here $A(t, T)$ and $B(t, T)$ are differently given dependent on the specific model of short rate [13]. For example, if the short rate follows the *Vasicek* model, that is, if the coefficients $a_1(t)$, $a_2(t)$, $s_r(t)$ in (1) are all constants (that is, $a_1(t) \equiv a_1, a_2(t) \equiv a_2, s_r(t) \equiv s_r$), then $B(t, T)$ and $A(t, T)$ are respectively given as follows:

$$B(t,T) = \frac{1-e^{-a_2(T-t)}}{a_2}, \quad A(t,T) = -\int_t^T \left[ a_2 B(u,T) - \frac{1}{2} s_r^2 B^2(u,T) \right] du. \tag{2.4}$$

See [13] for $B(t, T)$ and $A(t, T)$ in *Ho-Lee* model and *Hull-White* model.

  2) The firm value $V(t)$ follows a geometric Brown motion

$$dV(t) = (r_t - b)V(t)dt + s_V(t)V(t)dW_2(t)$$

under the risk neutral martingale measure and a standard Wiener process $W_2$ and $E(dW_1, dW_2) = \rho dt$. The firm continuously pays out dividend in rate $b \geq 0$ (constant) for a unit of firm value.

  3) Let $0 = T_0 < T_1 < \cdots < T_{N-1} < T_N = T$ and $T$ is the maturity of our corporate bond (debt) with face value $F$ (unit of currency). At time $T_i$ ($i = 1,\ldots, N-1$) bond holder receives the coupon of quantity $C_i \cdot Z(r, T_i\,; T)$ (unit of currency) from the firm and at time $T_N = T$ bond holder receives the face value $F$ and the last coupon $C_N$ (unit of currency). This means that the time $T$-value of the *sum of the face value and the coupons* of the bond is $F + \Sigma_{k=1}^N C_k$.

  4) The expected default occurs only at time $T_i$ when the equity of the firm is not enough to pay debt and coupon. If the expected default occurs, the bond holder receives $\delta \cdot V$ as *default recovery* and the equity holder gets nothing. Here $0 \leq \delta \leq 1$ is called a *fractional recovery rate* of firm value at default.

  5) The unexpected default can occur at any time. The unexpected default probability in the time interval $[t, t+\Delta t] \cap [T_i, T_{i+1}]$ is $\lambda_i \Delta t$ ($i = 0,\cdots, N-1$). Here the *default intensity* $\lambda_i$ is a constant. If the unexpected default occurs at time $t \in (T_i, T_{i+1})$, the bond holder receives $\min\{\delta \cdot V, (F + \Sigma_{k=i+1}^N C_k) \cdot Z(r, t\,; T)\}$ as default recovery and the equity holder gets nothing. Here the *reason* why the expected default recovery and unexpected recovery are *given* in *different* forms is to avoid the possibility of paying at time $t \in (T_i, T_{i+1})$ *more* than *the current price of risk free zero coupon bond* with the face value $F + \Sigma_{k=i+1}^N C_k$ as a default recovery when the unexpected default event occurs.

  6) In the subinterval $(T_i, T_{i+1}]$, the price of our corporate bond and the equity of the firm are given by a sufficiently smooth function $B_i(V, r, t)$ and $E_i(V, r, t)$ ($i = 0,\cdots, N-1$), respectively.

**2.2 Two Factors-Model for Equity and Expected Default Barriers**

  In order for us to understand the equity $E$, we derive the PDE for the equity using the $\Delta$-hedging technique [6, 13]. Construct a portfolio $\Pi = E - \Delta_1 B - \Delta_2 Z$. ($Z$ is the default free zero coupon bond and $B$ is the defaultable bond with a constant default intensity $\lambda$ and unexpected default recovery $R_{ud}$.) Choose $\Delta_1$ and $\Delta_2$ so that the portfolio $\Pi$ is risk-neutral in the time interval $[t, t+dt]$, that is,

$$d\Pi_t = r\Pi_t\, dt. \qquad (2.5)$$

If there is no default in the time interval $[t, t+dt]$, then

$$\begin{aligned}
d\Pi_t &= dE - \Delta_1 dB - \Delta_2 dZ = \\
&= \frac{\partial E}{\partial t}dt + \frac{1}{2}\left[\frac{\partial^2 E}{\partial V^2}(dV)^2 + 2\frac{\partial^2 E}{\partial V \partial r}dVdr + \frac{\partial^2 E}{\partial r^2}(dr)^2\right] + \frac{\partial E}{\partial V}dV + \frac{\partial E}{\partial r}dr \\
&\quad - \Delta_1\left\{\frac{\partial B}{\partial t}dt + \frac{1}{2}\left[\frac{\partial^2 B}{\partial V^2}(dV)^2 + 2\frac{\partial^2 B}{\partial V \partial r}dVdr + \frac{\partial^2 B}{\partial r^2}(dr)^2\right] + \frac{\partial B}{\partial V}dV + \frac{\partial B}{\partial r}dr\right\} \\
&\quad - \Delta_2\left\{\frac{\partial Z}{\partial t}dt + \frac{1}{2}\frac{\partial^2 Z}{\partial r^2}(dr)^2 + \frac{\partial Z}{\partial r}dr\right\} \\
&= \frac{\partial E}{\partial t}dt + \frac{1}{2}\left[\frac{\partial^2 E}{\partial V^2}(dV)^2 + 2\frac{\partial^2 E}{\partial V \partial r}dVdr + \frac{\partial^2 E}{\partial r^2}(dr)^2\right] - \Delta_2\left\{\frac{\partial Z}{\partial t}dt + \frac{1}{2}\frac{\partial^2 Z}{\partial r^2}(dr)^2\right\} \\
&\quad - \Delta_1\left\{\frac{\partial B}{\partial t}dt + \frac{1}{2}\left[\frac{\partial^2 B}{\partial V^2}(dV)^2 + 2\frac{\partial^2 B}{\partial V \partial r}dVdr + \frac{\partial^2 B}{\partial r^2}(dr)^2\right]\right\} \\
&\quad + \left(\frac{\partial E}{\partial V} - \Delta_1\frac{\partial B}{\partial V}\right)dV + \left(\frac{\partial E}{\partial r} - \Delta_1\frac{\partial B}{\partial r} - \Delta_2\frac{\partial Z}{\partial r}\right)dr.
\end{aligned}$$

Let

$$\Delta_1 = \frac{\partial E}{\partial V}\bigg/\frac{\partial B}{\partial V},\quad \Delta_2 = \left(\frac{\partial E}{\partial r} - \Delta_1\frac{\partial B}{\partial r}\right)\bigg/\frac{\partial Z}{\partial r} = \left(\frac{\partial E}{\partial r} - \frac{\partial E}{\partial V}\frac{\partial B}{\partial r}\bigg/\frac{\partial B}{\partial V}\right)\bigg/\frac{\partial Z}{\partial r}. \qquad (2.6)$$

If we consider the assumptions 1) and 2) and neglect higher order infinitesimal of $dt$, then

$$d\Pi_t = \left\{\frac{\partial E}{\partial t} + \frac{1}{2}\left(s_V^2 V^2 \frac{\partial^2 E}{\partial V^2} + 2\rho\, s_V s_r V \frac{\partial^2 E}{\partial V \partial r} + s_r^2 \frac{\partial^2 E}{\partial r^2}\right) - \Delta_2\left(\frac{\partial Z}{\partial t} + \frac{1}{2}s_r^2\frac{\partial^2 Z}{\partial r^2}\right)\right. \\
\left. - \Delta_1\left[\frac{\partial B}{\partial t} + \frac{1}{2}\left(s_V^2 V^2 \frac{\partial^2 B}{\partial V^2} + 2\rho\, s_V s_r V \frac{\partial^2 B}{\partial V \partial r} + s_r^2 \frac{\partial^2 B}{\partial r^2}\right)\right]\right\}dt. \qquad (2.7)$$

If the unexpected default occurs in the time interval $[t, t+dt]$, then $E_{t+dt}$ becomes 0 and the bond holders receive $R_{ud}$, and therefore we have

$$d\Pi_t = -E - \Delta_1(R_{ud} - B) - \Delta_2\left[\left(\frac{\partial Z}{\partial t} + \frac{1}{2}s_r^2\frac{\partial^2 Z}{\partial r^2}\right)dt + \frac{\partial Z}{\partial r}dr\right]. \qquad (2.8)$$

Note that the probability of default in the time interval $[t, t+dt]$ is $\lambda dt$. Multiply $1 - \lambda dt$ to (2.7) and $\lambda dt$ to (2.8), and then add them together and neglect higher order infinitesimal terms of $dt$, then we have

$$d\Pi_t = \left\{ \frac{\partial E}{\partial t} + \frac{1}{2}\left( s_V^2 V^2 \frac{\partial^2 E}{\partial V^2} + 2\rho\, s_V s_r V \frac{\partial^2 E}{\partial V \partial r} + s_r^2 \frac{\partial^2 E}{\partial r^2} \right) - \Delta_2 \left( \frac{\partial Z}{\partial t} + \frac{1}{2} s_r^2 \frac{\partial^2 Z}{\partial r^2} \right) \right.$$
$$\left. - \Delta_1 \left[ \frac{\partial B}{\partial t} + \frac{1}{2}\left( s_V^2 V^2 \frac{\partial^2 B}{\partial V^2} + 2\rho\, s_V s_r V \frac{\partial^2 B}{\partial V \partial r} + s_r^2 \frac{\partial^2 B}{\partial r^2} \right) \right] - \lambda E - \lambda \Delta_1 (R_{ud} - B)) \right\} dt$$

If we substitute this expression into (2.5) and remove $dt$, then we have

$$\frac{\partial E}{\partial t} + \frac{1}{2}\left( s_V^2 V^2 \frac{\partial^2 E}{\partial V^2} + 2\rho\, s_V s_r V \frac{\partial^2 E}{\partial V \partial r} + s_r^2 \frac{\partial^2 E}{\partial r^2} \right) - (r+\lambda)E - \Delta_2 \left( \frac{\partial Z}{\partial t} + \frac{1}{2} s_r^2 \frac{\partial^2 Z}{\partial r^2} - rZ \right)$$
$$- \Delta_1 \left[ \frac{\partial B}{\partial t} + \frac{1}{2}\left( s_V^2 V^2 \frac{\partial^2 B}{\partial V^2} + 2\rho\, s_V s_r V \frac{\partial^2 B}{\partial V \partial r} + s_r^2 \frac{\partial^2 B}{\partial r^2} \right) - (r+\lambda)B + \lambda R_{ud} \right] = 0.$$

If we consider the equation (2.2) for $Z$ and the equation of the defaultable bond with a constant default intensity $\lambda$ and unexpected default recovery $R_{ud}$ (see [13]) :

$$\frac{\partial B}{\partial t} + \frac{1}{2}\left[ s_V^2 V^2 \frac{\partial^2 B}{\partial V^2} + 2\rho\, s_V s_r V \frac{\partial^2 B}{\partial V \partial r} + s_r^2 \frac{\partial^2 B}{\partial r^2} \right] + (r-b)V \frac{\partial B}{\partial V} + a_r \frac{\partial B}{\partial r} - (r+\lambda)B + \lambda R_{ud} = 0, \quad (2.9)$$

then, we have

$$\frac{\partial E}{\partial t} + \frac{1}{2}\left( s_V^2 V^2 \frac{\partial^2 E}{\partial V^2} + 2\rho\, s_V s_r V \frac{\partial^2 E}{\partial V \partial r} + s_r^2 \frac{\partial^2 E}{\partial r^2} \right) - (r+\lambda)E + \Delta_2 a_r \frac{\partial Z}{\partial r} + \Delta_1 \left[ (r-b)V \frac{\partial B}{\partial V} + a_r \frac{\partial B}{\partial r} \right] = 0.$$

If we consider (2.6), then

$$\frac{\partial E}{\partial t} + \frac{1}{2}\left( s_V^2 V^2 \frac{\partial^2 E}{\partial V^2} + 2\rho\, s_V s_r V \frac{\partial^2 E}{\partial V \partial r} + s_r^2 \frac{\partial^2 E}{\partial r^2} \right) + (r-b)V \frac{\partial E}{\partial V} + a_r \frac{\partial E}{\partial r} - (r+\lambda)E = 0.$$

This means that the *equity E* (when the firm has constant default intensity $\lambda$) satisfies the same PDE for *defaultable security* with 0-unexpected default recovery.

Now we derive the mathematical model for the equity under the assumptions 1) ~ 6). From the above PDE of the equity and the above assumption 5), 6) the equity price $E_i$ satisfies the following PDE in every subinterval $(T_i, T_{i+1})$ ($i = 0, \cdots, N-1$):

$$\frac{\partial E_i}{\partial t} + \frac{1}{2}\left( s_V^2 V^2 \frac{\partial^2 E_i}{\partial V^2} + 2\rho\, s_V s_r V \frac{\partial^2 E_i}{\partial V \partial r} + s_r^2 \frac{\partial^2 E_i}{\partial r^2} \right) + (r-b)V \frac{\partial E_i}{\partial V} + a_r \frac{\partial E_i}{\partial r} - (r+\lambda_i)E_i = 0. \quad (2.10)$$

From the assumption 3) and 4) we have:

$$E_{N-1}(V, r, T_N) = (V - F - C_N) \cdot 1\{V \geq F + C_N\},$$
$$E_i(V, r, T_{i+1}) = [E_{i+1}(V, r, T_{i+1}) - C_{i+1} Z(r, T_{i+1}; T_N)] \cdot 1\{E_{i+1}(V, r, T_{i+1}) \geq C_{i+1} Z(r, T_{i+1}; T_N)\}, \quad (2.11)$$
$$i = 0, \cdots, N-2.$$

The problem (2.10) and (2.11) is just the mathematical model for the equity.

**Remark 1**. The equation (2.10) is the same type with (3.7) in [11] but simpler than (3.7) (that is, (2.10) is homogeneous). In the first expression of (2.11) the default barrier is *explicitly* shown.

But in the second expression, the default conditions don't show the default barrier explicitly. So in this stage, it is still not clear to find the similarity of this problem with the problem of [10] but through more careful consideration we can use the method of [10] to get the pricing formula of our equity (see the proof in the section 5).

When $B(t,T)$ is given in (2.4), let denote

$$S_x^2(t) = S_x^2(t;T) = s_V^2(t) + 2\rho s_V(t) \cdot s_r(t) \cdot B(t,T) + (s_r(t)B(t,T))^2$$
$$K_N = F + C_N; \bar{c}_N = F + C_N; \bar{c}_i = C_i, i = 1, \cdots, N-1;$$
$$\Delta T_i = T_{i+1} - T_i, i = 0, \cdots, N-1.$$
(2.12)

**Remark 2.** $S_x(t)$ is the volatility of the relative price of the firm value $x = V/Z$, $\bar{c}_i$ is the time $T_N$-value of the payoff to bondholders at time $T_i$ ($i = 1, \ldots, N$) and $K_N$ denotes the default barrier at time $T_N$ as in [10].

**Theorem 1.** (Equity Price) *The solutions of* (2.10) *and* (2.11) *are provided as follows:*

$$E_i(V, r, t) = Z(r,t;T_N) \cdot e_i(V/Z(r,t;T_N), t), T_i < t \le T_{i+1}, \quad i = 0, \cdots, N-1.$$
(2.13)

*Here*

$$e_i(x, t) = e^{-\lambda_i(T_{i+1}-t)} \left\{ e^{-\sum_{k=i+1}^{N-1} \lambda_k \Delta T_k} A_{K_{i+1}\cdots K_N}^{+\cdots +}(x, t; T_{i+1}, \cdots, T_N; 0, b, S_x(\cdot)) \right.$$
$$\left. - \sum_{m=i}^{N-1} \bar{c}_{m+1} e^{-\sum_{k=i+1}^{m} \lambda_k \Delta T_k} B_{K_{i+1}\cdots K_{m+1}}^{+\cdots +}(x, t; T_{i+1}, \cdots, T_{m+1}; 0, b, S_x(\cdot)) \right\}.$$
(2.14)

*Here* $B_{K_1\cdots K_m}^{+\cdots +}(x,t;T_1,\cdots,T_m;0,b,S_x(\cdot))$ *and* $A_{K_1\cdots K_{m-1}K_m}^{+\cdots + +}(x,t;T_1,\cdots,T_{m-1},T_m;0,b,S_x(\cdot))$ *are the prices of m-th order bond and asset binaries with 0-risk free rate, b-dividend rate and* $S_x(t)$-*volatility (see the theorem 1 of [10]) and* $K_i (i=1,\cdots,N-1)$ *is the unique root of the equation* $e_i(x,T_i) = C_i$. *Furthermore* $e_i(x, t)$ *is x-increasing and x-downward convex and* $0 < \partial_x e_i < 1$.

**Remark 3.** 1) The theorem 1 gives us the expected default barrier $K_i$ at time $T_i$ ($i = 1, \ldots, N-1$). That is, if $V < K_i \cdot Z(r, T_i ; T)$ at time $T_i$, then the expected default occurs. 2) Using multi-variate normal distribution functions, (2.13) are represented in terms of the debt $F$, the coupons $C_i$ and the firm value $V$ as follows:

$$E_i(V, r, t) = V \cdot e^{-(\lambda_i+b)(T_{i+1}-t) - \sum_{k=i+1}^{N-1}(\lambda_k+b)\Delta T_k} N_{N-i}(d_{i+1}^+(t), \cdots, d_N^+(t); A_{i+1, N}(t))$$
$$- (F + C_N) \cdot Z(r,t;T_N) e^{-\lambda_i(T_{i+1}-t) - \sum_{k=i+1}^{N-1} \lambda_k \Delta T_k} N_{N-i}(d_{i+1}^-(t), \cdots, d_N^-(t); A_{i+1, N}(t)) -$$
$$- \sum_{m=i}^{N-2} C_{m+1} Z(r,t;T_N) e^{-\lambda_i(T_{i+1}-t) - \sum_{k=i+1}^{m} \lambda_k \Delta T_k} N_{m+1-i}(d_{i+1}^-(t), \cdots, d_{m+1}^-(t); A_{i+1, m+1}(t)). (2.15)$$

Here $N_m(a_1, \cdots, a_m ; A)$ ( the *cumulative distribution function* of *m*-variate normal distribution with *zero mean* vector and a *covariance* matrix $A^{-1}$), $d_i^\pm(t)$ and $(A_{k,m}(t))^{-1} = (r_{ij}(t))_{i,j=k}^m$ *are given by*:

$$N_m(a_1,\cdots,a_m \,;A) = \int_{-\infty}^{a_1}\cdots\int_{-\infty}^{a_m} \frac{1}{(\sqrt{2\pi})^m} \sqrt{\det A}\, \exp(-\frac{1}{2} y^\perp Ay)dy,$$

$$d_i^\pm(t) = \left(\int_t^{T_i} S_x^2(u)du\right)^{-1/2}\left(\ln\frac{V}{K_i Z(r,t\,;T_N)} - b(T_i - t) \pm \frac{1}{2}\int_t^{T_i} S_x^2(u)du\right), i = 1,\cdots,N-1,$$

$$d_N^\pm(t) = \left(\int_t^{T_N} S_x^2(u)du\right)^{-1/2}\left(\ln\frac{V}{(F+C_N)Z(r,t\,;T_N)} - b(T_N - t) \pm \frac{1}{2}\int_t^{T_N} S_x^2(u)du\right),$$

$$r_{i\,j}(t) = \sqrt{\int_t^{T_i} S_x^2(u)du\Big/\int_t^{T_j} S_x^2(u)du}\,, \ r_{j\,i}(t) = r_{i\,j}(t),\ i \le j\,(i,\,j = k,\cdots,m). \qquad (2.16)$$

3) If $b = 0$ and $\lambda_k = 0$ ($i = 0,\ldots, N-1$), then our pricing formula (2.15) nearly coincides with the formula (2) of [1, at page 751] and the only difference comes from the fact that $k$-th coupon is provided as a discounted value of the maturity-value in our model.

## 2.3 Two Factors-Model Pricing Formulae of the Defaultable Discrete Coupon Bond

In this subsection we derive the representation of the price $B_i(V, r, t)$ of the defaultable discrete coupon bond in the interval $(T_i, T_{i+1}]$ ($i = 0,\ldots, N-1$). In this subsection we *neglect* the effect of the *taxation*. We use the notation of (2.12) and the following notation

$$\Phi_i = F + \Sigma_{k=i+1}^N C_k \,. \qquad (2.17)$$

That is, $\Phi$ is the time $T_N$-value of the sum of the face value and all coupons of the bond.

Now we consider the defaultable discrete coupon bond under the assumptions 1) ~ 6) in the subsection 2.1. From (2.9) and the assumption 5) and 6) we can know that our bond price $B_i$ satisfies the following PDE in every subinterval $(T_i, T_{i+1})$ ($i = 0,\cdots, N-1$):

$$\frac{\partial B_i}{\partial t} + \frac{1}{2}\left[s_V^2 V^2 \frac{\partial^2 B_i}{\partial V^2} + 2\rho s_V s_r V \frac{\partial^2 B_i}{\partial V \partial r} + s_r^2 \frac{\partial^2 B_i}{\partial r^2}\right] + (r-b)V\frac{\partial B_i}{\partial V} + a_r \frac{\partial B_i}{\partial r} - (r + \lambda_i)B_i \qquad (2.18)$$
$$+ \lambda_i \min\{\delta \cdot V,\ \Phi_i \cdot Z(r,t\,;T)\} = 0, \qquad T_i < t < T_{i+1},\ V > 0.$$

In the theorem 1, we have calculated the expected default barrier $K_i$ ($i = 1,\ldots, N$) (see the remark 3). Thus from the assumptions 3) and 4) we have the following terminal value conditions:

$$B_{N-1}(V,r,T_N) = \bar{c}_N \cdot 1\{V \ge K_N\} + \delta V \cdot 1\{V < K_N\},$$
$$B_i(V,r,T_{i+1}) = [B_{i+1}(V,r,T_{i+1}) + \bar{c}_{i+1} Z(r,T_{i+1};T)] \cdot 1\{V \ge K_{i+1}Z(r,T_{i+1};T)\} + \qquad (2.19)$$
$$+ \delta V \cdot 1\{V < K_{i+1}Z(r,T_{i+1};T)\},\ V > 0, \quad i = 0,\cdots, N-2.$$

The problem (2.18) and (2.19) is just the pricing model of our defaultable discrete coupon bond.

**Remark 4**. In our model (2.18) and (2.19) the consideration of *unexpected default risk* and dividend of firm value is added to the model on defaultable discrete coupon bond of [1]. Another difference from [1]'s approach is that the bond is considered as a derivative of risk free rate $r$ (but not default free zero coupon bond) and the coupons prior to maturity is provided as a discounted value of the maturity-value. The difference from the model (5.5) ~ (5.7) of [9] is that here we consider the different unexpected default intensity $\lambda_i$ in the every subinterval $(T_i, T_{i+1}]$ and the expected default barriers are calculated from the equity price like [1].

The equation (2.18) and (2.19) is just the same type with (3.7) in [10]. So we can use the method of [10] or [11] to get the following pricing formula.

**Theorem 2.** (Discrete Coupon Bond Price) *The solution of* (2.18) *and* (2.19) *is given by:*

$$B_i(V,r,t) = Z(r,t;T_N) \cdot u_i(V/Z(r,t;T_N),t), T_i < t < T_{i+1}, \ i=0,\cdots,N-1. \qquad (2.20)$$

*Here* $u_i(x,t), T_i < t \leq T_{i+1}, \ i=0,\cdots,N-1$ *are the solutions to the problems*

$$\frac{\partial u_i}{\partial t} + \frac{1}{2} S_x^2(t) x^2 \frac{\partial^2 u_i}{\partial x^2} - bx \frac{\partial u_i}{\partial x} - \lambda u_i + \lambda \min\{\Phi_i, \delta x\} = 0, \ x>0, \ T_i<t<T_{i+1}, i=0,\cdots,N-1,$$

$$u_{N-1}(x,T_N) = \bar{c}_N \cdot 1\{x > K_N\} + \delta x \cdot 1\{x \leq K_N\}, \qquad (2.21)$$

$$u_i(x,T_{i+1}) = [u_{i+1}(x,T_{i+1}) + \bar{c}_{i+1}] \cdot 1\{x \geq K_{i+1}\} + \delta x \cdot 1\{x < K_{i+1}\}, i = \overline{0, N-2}.$$

*and provided as follows:*

$$u_i(x,t) = e^{-\lambda_i(T_{i+1}-t)} \left\{ \sum_{m=i}^{N-1} e^{-\sum_{k=i+1}^{m} \lambda_k \Delta T_k} \left[ \bar{c}_{m+1} B^{+\cdots+}_{K_{i+1}\cdots K_{m+1}}(x,t;T_{i+1},\cdots,T_{m+1};0,b,S_x(\cdot)) \right.\right.$$

$$\left. + \delta \cdot A^{+\cdots+-}_{K_{i+1}\cdots K_m K_{m+1}}(x,t;T_{i+1},\cdots,T_m,T_{m+1};0,b,S_x(\cdot)) \right]$$

$$+ \sum_{m=i+1}^{N-1} \lambda_m e^{-\sum_{k=i+1}^{m-1} \lambda_k \Delta T_k} \int_{T_m}^{T_{m+1}} e^{-\lambda_m(\tau-T_m)} \left[ \Phi_m \cdot B^{+\cdots++}_{K_{i+1}\cdots K_m \Phi_m/\delta}(x,t;T_{i+1},\cdots,T_m,\tau;0,b,S_x(\cdot)) \right.$$

$$\left.\left. + \delta \cdot A^{+\cdots+-}_{K_{i+1}\cdots K_m \Phi_m/\delta}(x,t;T_{i+1},\cdots,T_m,\tau;0,b,S_x(\cdot)) \right] d\tau \right\}$$

$$+ \lambda_i \int_t^{T_{i+1}} e^{-\lambda_i(\tau-t)} \left[ \Phi_i B^{+}_{\Phi_i/\delta}(x,t;\tau;0,b,S_x(\cdot)) + \delta \cdot A^{-}_{\Phi_i/\delta}(x,t;\tau;0,b,S_x(\cdot)) \right] d\tau, T_i < t \leq T_{i+1}, x>0.$$

*Here* $B^{+\cdots+}_{K_1\cdots K_m}$, $A^{+\cdots+-}_{K_1\cdots K_{m-1}K_m}$, $\bar{c}_i, K_i(i=1,\cdots,N)$ *are the same with the theorem* 1. *In particular the initial price of the bond is given by*

$$B_0 = B_0(V_0,r_0,0) = $$

$$= Z_0 \sum_{m=0}^{N-1} e^{-\sum_{k=0}^{m-1} \lambda_k \Delta T_k} \left\{ e^{-\lambda_m \Delta T_m} \left[ \bar{c}_{m+1} B^{+\cdots+}_{K_1\cdots K_{m+1}}(V_0/Z_0,0;T_1,\cdots,T_{m+1};0,b,S_x(\cdot)) \right.\right.$$

$$\left. + \delta \cdot A^{+\cdots+-}_{K_1\cdots K_m K_{m+1}}(V_0/Z_0,0;T_1,\cdots,T_m,T_{m+1};0,b,S_x(\cdot)) \right] \qquad (2.22)$$

$$+ \lambda_m \int_{T_m}^{T_{m+1}} e^{-\lambda_m(\tau-T_m)} \left[ \Phi_m \cdot B^{+\cdots++}_{K_1\cdots K_m \Phi_m/\delta}(V_0/Z_0,0;T_1,\cdots,T_m,\tau;0,b,S_x(\cdot)) \right.$$

$$\left.\left. + \delta \cdot A^{+\cdots+-}_{K_1\cdots K_m \Phi_m/\delta}(V_0/Z_0,0;T_1,\cdots,T_m,\tau;0,b,S_x(\cdot)) \right] d\tau \right\}.$$

*Here* $Z_0 = Z(r_0,0;T)$.

**Remark 5.** 1) The proof of theorem 2 is omitted since it is very similar with the theorem 1 of [11] or the theorem 3 of [10]. 2) The equation (2.21) of the relative price $u_i(x,t)$ is an inhomogenous Black-Scholes equation with discontinuous terminal value. Thus using the results of [9], we can investigate such properties of $u_i(x,t)$ as monotonicity, boundedness or gradient estimate and so on. For example, $u_{N-1}(x,t)$ is $x$ - increasing. $u_{N-2}(x,t)$ is $x$ - increasing if $u_{N-1}(K_{N-1},T_{N-1}) + \bar{c}_{N-1} \geq \delta \cdot K_{N-1}$.

Let denote the ***leverage ratio*** by $L=F/V_0$ and the $k$-th ***coupon rate*** by $c_k = C_k/F$ ($k = 1,\ldots,$

*N*). Then we have the following *representation* of the *initial price* of the our *defaultable discrete coupon bond* in terms of *leverage ratio, coupon rates, default recovery rate* and *initial price of the default free zero coupon bond*.

**Corollary 1.** *Under the assumption of theorem* 1, *the initial price of the bond can be represented as follows*:

$$B_0 = B_0(F, c_1, \cdots, c_N; \delta, \lambda_0, \cdots, \lambda_{N-1}; Z_0, L) =$$

$$= FZ_0 \left[ e^{-\sum_{k=0}^{N-1} \lambda_k \Delta T_k} N_N(d_1^-, \cdots, d_{N-1}^-, d_N^-; A_N) + \sum_{m=0}^{N-1} c_{m+1} e^{-\sum_{k=0}^{m} \lambda_k \Delta T_k} N_{m+1}(d_1^-, \cdots, d_{m+1}^-; A_{m+1}) \right]$$

$$+ \delta \frac{F}{L} \sum_{m=0}^{N-1} e^{-\sum_{k=0}^{m}(\lambda_k + b)\Delta T_k} N_{m+1}(d_1^+, \cdots, d_m^+, -d_{m+1}^+; A_{m+1}^-)$$

$$+ \sum_{m=0}^{N-1} \lambda_m \int_{T_m}^{T_{m+1}} \left[ \delta \frac{F}{L} e^{-\sum_{k=0}^{m-1}(\lambda_k + b)\Delta T_k} e^{-(\lambda_m + b)(\tau - T_m)} N_{m+1}(d_1^+, \cdots, d_m^+, -\tilde{d}_{m+1}^+(\tau, \delta); \tilde{A}_{m+1}^-(\tau)) + \right.$$

$$\left. + F(1 + \Sigma_{k=m+1}^N c_k) Z_0 e^{-\sum_{k=0}^{m-1} \lambda_k \Delta T_k} e^{-\lambda_m(\tau - T_m)} N_{m+1}(d_1^-, \cdots, d_m^-, \tilde{d}_{m+1}^-(\tau, \delta); \tilde{A}_{m+1}^-(\tau)) \right] d\tau. \quad (2.23)$$

*Here* $N_m(a_1, \cdots, a_m; A)$ *is given by* (2.16), $d_i^\pm$ *and* $(A_m)^{-1} = (r_{ij})_{i,j=1}^m$ *are provided by*

$$d_i^\pm = d_i^\pm(0) = \left( \int_0^{T_i} S_x^2(u) du \right)^{-1/2} \left( \ln \frac{F}{LZ_0 K_i} - bT_i \pm \frac{1}{2} \int_0^{T_i} S_x^2(u) du \right), i = 1, \cdots, N-1;$$

$$d_N^\pm = d_N^\pm(0) = \left( \int_0^{T_N} S_x^2(u) du \right)^{-1/2} \left( \ln \frac{1}{LZ_0(1 + c_N)} - bT_N \pm \frac{1}{2} \int_0^{T_N} S_x^2(u) du \right);$$

$$\tilde{d}_i^\pm(\tau, \delta) = \left( \int_0^\tau S_x^2(u) du \right)^{-1/2} \left( \ln \frac{\delta}{LZ_0(1 + \Sigma_{k=i}^N c_k)} - b\tau \pm \frac{1}{2} \int_0^\tau S_x^2(u) du \right), T_{i-1} \leq \tau < T_i; i = 1, \cdots, N,$$

$$r_{ij} = \sqrt{\int_0^{T_i} S_x^2(t) dt \Big/ \int_0^{T_j} S_x^2(t) dt}, \; r_{ji} = r_{ij}, \; i \leq j \; (i, j = 1, \cdots, m). \quad (2.24)$$

$(\tilde{A}_m(\tau))^{-1} = (\tilde{r}_{ij}(\tau))_{i,j=1}^m$ *is the matrix whose m-th row and column are given by*

$$\tilde{r}_{im}(\tau) = \sqrt{\int_0^{T_i} S_X^2(t) dt \Big/ \int_0^\tau S_X^2(t) dt}, \; \tilde{r}_{mi}(\tau) = \tilde{r}_{im}(\tau), \; i < m \; (i = 1, \cdots, m-1) \quad (2.25)$$

*and other elements coincide with those of* $(A_m)^{-1}$. *The matrices* $(A_m^-)^{-1} = (r_{ij}^-)_{i,j=1}^m$ *and* $(\tilde{A}_m^-(\tau))^{-1}$ $= (\tilde{r}_{ij}^-(\tau))_{i,j=1}^m$ *have such m-th rows and columns that*

$$r_{im}^- = -r_{im}, \; r_{mi}^- = -r_{mi}; \; \tilde{r}_{im}^-(\tau) = -\tilde{r}_{im}(\tau), \; \tilde{r}_{mi}^-(\tau) = -\tilde{r}_{mi}(\tau), \; i < m \; (i = 1, \cdots, m-1)$$

*and other elements coincide with those of* $(A_m)^{-1}$ *and* $(\tilde{A}_m(\tau))^{-1}$, *respectively* .

**Remark 6.** If $\lambda_k = \lambda$ ($i = 0, \ldots, N-1$), then our pricing formula (2.23) coincides with the formula (5.34) of [9] but the calculated default barrier $K_i$, $i = 1, \ldots, N-1$ may be different. If $b = 0$ and $\lambda_k = 0$ ($i = 0,\ldots, N-1$), then our pricing formula (2.23) nearly coincides with the formula (5) of [1, at page 752] and the only difference comes from the fact that *k*-th coupon is provided as a discounted value of the maturity-value in our model. If $b = 0$, $C_k = 0$ and $\lambda_k = 0$ ($i = 0,\ldots, N-1$), then the formula (2.23) coincides with the formula (5) of [1, at page 752] when $C_k = 0$ ($i = 0,\ldots, N-1$) which is a known formula for defaultable zero coupon bond that generalizes Merton (1974) [7].

*Note* that in the formula (2.23), the first term reflects the *survival probability*, more exactly speaking, $e^{-\sum_{k=0}^{m}\lambda_k \Delta T_k} N_{m+1}(d_1^-,\cdots,d_{m+1}^-; A_{m+1})$ can be interpreted as the survival (no default) probability at time $T_{m+1}$ under the condition that expected or unexpected default has not occurred in $[0, T_{m+1})$; the second term is the *expected default premium* (more exactly speaking, the default premium when unexpected default recovery is zero) and the last term means the *unexpected default premium* (more exactly speaking, the default premium coming from only the possibility of unexpected default). $e^{-\sum_{k=0}^{m}(\lambda_k+b)\Delta T_k} N_{m+1}(d_1^+,\cdots,d_m^+, -d_{m+1}^+; A_{m+1}^-)$ can be interpreted as the expected default probability at time $T_{m+1}$ under the condition that expected or unexpected default has not occurred in $[0, T_{m+1})$ and other terms have similar explanations.

In what follows, we use the following notation for simplicity:

$$G_N^+ = G_N^+(\lambda_0,\cdots,\lambda_{N-1}; b) = e^{-\sum_{k=0}^{N-1}(\lambda_k+b)\Delta T_k} N_N(d_1^+,\cdots,d_N^+; A_N),$$

$$G_{m+1}^- = G_{m+1}^-(\lambda_0,\cdots,\lambda_m) = e^{-\sum_{k=0}^{m}\lambda_k \Delta T_k} N_{m+1}(d_1^-,\cdots,d_{m+1}^-; A_{m+1})$$

$$g_{m+1}^-(\tau) = g_{m+1}^-(\tau; \delta, \lambda_0,\cdots,\lambda_m) = e^{-\sum_{k=0}^{m-1}\lambda_k \Delta T_k} e^{-\lambda_m(\tau - T_m)} N_{m+1}(d_1^-,\cdots,d_m^-, \tilde{d}_{m+1}^-(\tau,\delta); \tilde{A}_{m+1}^-(\tau)), \quad (2.26)$$

$$\tilde{G}_{m+1} = \tilde{G}_{m+1}(\lambda_0,\cdots,\lambda_m; b) = e^{-\sum_{k=0}^{m}(\lambda_k+b)\Delta T_k} N_m(d_1^+,\cdots,d_m^+, -d_{m+1}^+; A_{m+1}^-),$$

$$\tilde{g}_{m+1}(\tau) = \tilde{g}_{m+1}(\tau; \delta, \lambda_0,\cdots,\lambda_m; b) =$$
$$= e^{-(\lambda_m+b)(\tau-T_m)-\sum_{k=0}^{m-1}(\lambda_k+b)\Delta T_k} N_{m+1}(d_1^+,\cdots,d_m^+, -\tilde{d}_{m+1}^+(\tau,\delta); \tilde{A}_{m+1}^-(\tau)), \quad m=0,\cdots,N-1.$$

Then from (2.15) and (2.23) we can write as follows:

$$E_0(V_0, r, 0) = V_0 G_N^+ - Z_0\left(FG_N^- + \sum_{m=0}^{N-1} C_{m+1} G_{m+1}^-\right),$$

$$B_0(V_0, r, 0) = Z_0\left[FG_N^- + \sum_{m=0}^{N-1} C_{m+1} G_{m+1}^- + (F + \Sigma_{k=1}^N C_k)\sum_{m=0}^{N-1} \lambda_m \int_{T_m}^{T_{m+1}} g_{m+1}^-(\tau)d\tau\right] +$$
$$+ \delta V_0 \sum_{m=0}^{N-1}\left(\tilde{G}_{m+1} + \lambda_m \int_{T_m}^{T_{m+1}} \tilde{g}_{m+1}(\tau)d\tau\right). \quad (2.27)$$

If we take the sum of the above expressions, we have

$$E_0 + B_0 = V_0 G_N^+ + \delta V_0 \sum_{m=0}^{N-1}\left(\tilde{G}_{m+1} + \lambda_m \int_{T_m}^{T_{m+1}} \tilde{g}_{m+1}(\tau)d\tau\right) + (F + \Sigma_{k=1}^N C_k)\sum_{m=0}^{N-1} \lambda_m \int_{T_m}^{T_{m+1}} g_{m+1}^-(\tau)d\tau.$$

Therefore we have

$$V_0 = E_0 + B_0 + V_0 \left[ 1 - G_N^+ - \delta \sum_{m=0}^{N-1} \left( \tilde{G}_{m+1} + \lambda_m \int_{T_m}^{T_{m+1}} \tilde{g}_{m+1}(\tau) d\tau \right) \right]$$
$$- (F + \Sigma_{k=1}^{N} C_k) \sum_{m=0}^{N-1} \lambda_m \int_{T_m}^{T_{m+1}} g_{m+1}^-(\tau) d\tau.$$

This shows that the Modigliani-Miller theorem holds (that is, $V = Equity + Debt$) when $\delta = 1$ and $\lambda_k = b = 0$. Here we considered the following fact [1]:

$$1 - N_N(d_1^+, \cdots, d_N^+; A_N) = \sum_{m=0}^{N-1} N_{m+1}(d_1^+, \cdots, d_m^+, -d_{m+1}^+; A_{m+1}^-).$$

In the case with possibility of default, it is modified as follows [1]:

$$V = Equity + Debt + Default\ Costs\ (bankruptcy\ costs).$$

From this fact, we have the representation of *bankruptcy costs*.

**Corollary 2.** (Bankruptcy Cost) *The current value of bankruptcy cost is as follows:*

$$V_0 - V_0 G_N^+ - \sum_{m=0}^{N-1} \left\{ V_0 \delta \tilde{G}_{m+1} + \lambda_m \int_{T_m}^{T_{m+1}} [V_0 \delta \tilde{g}_{m+1}(\tau) + (F + \Sigma_{k=1}^{N} C_k) g_{m+1}^-(\tau)] d\tau \right\}. \qquad (2.28)$$

**Remark 7.** In the formula (2.28), let $\lambda_k = b = 0$, then we have the formula (6) of [1, at page 752].

## 3. Duration

In this section we study the problem of duration for defaultable discrete coupon bond under the united model of structural and reduced form approaches we developed in the previous section. According to [1], when $B(V,r,t)$ is bond price, we use the following definition for *duration* with respect to the short rate

$$D(V,r,t) = -\frac{1}{B(V,r,t)} \partial_r B(V,r,t). \qquad (3.1)$$

For example, the *duration* $D_z(t, T)$ of *default free zero coupon bond* $Z(r, t ; T)$ under the assumption 1) in the section 2 is just the $B(t, T)$ in the formula (2.3) and it is provided by (2.4) under the *Vasicek* model, that is,

$$D_Z(t,T) = -\frac{1}{Z(r,t;T)} \partial_r Z(r,t;T) = B(t,T) = \frac{1 - e^{-a_2(T-t)}}{a_2}. \qquad (3.2)$$

The *duration* of the *default free discrete coupon bond*, the holder of which receives the coupon of quantity $C_i \cdot Z(r, T_i ; T)$ (unit of currency) at time $T_i$ ($i = 1,\ldots, N-1$) and receives the face value $F$ and the last coupon $C_N$ (unit of currency) at time $T_N = T$, is given by (3.2) under the *Vasicek* model, too because its time *t*-price is just the same with $(F + \Sigma_{k=1}^{N} C_k) Z(r,t;T)$.

Now let calculate the duration of our defaultable discrete coupon bond. If we let in (2.27)

$$f_1 = FG_N^- + \sum_{m=0}^{N-1} C_{m+1} G_{m+1}^- + (F + \Sigma_{k=1}^{N} C_k) \sum_{m=0}^{N-1} \lambda_m \int_{T_m}^{T_{m+1}} g_{m+1}^-(\tau) d\tau,$$

$$f_2 = \sum_{m=0}^{N-1} \left( \tilde{G}_{m+1} + \lambda_m \int_{T_m}^{T_{m+1}} \tilde{g}_{m+1}(\tau) d\tau \right), \tag{3.3}$$

then $f_1, f_2 > 0$ and first time price of our bond is written as follows:

$$B_0 = B_0(V_0, Z_0(r,0), 0) = Z_0 f_1 + \delta V_0 f_2. \tag{3.4}$$

Thus we have $\partial_r B_0 = \partial_r Z_0 \cdot f_1 + Z_0 \cdot \partial_r f_1 + \delta V_0 \cdot \partial_r f_2$. We use the lemma on derivatives of multi-variate normal distribution functions (the lemma 1 in the section 5) and

$$\frac{\partial}{\partial r} d_i^\pm(0) = \left( \int_0^{T_i} S_x^2(u) du \right)^{-1/2} \left( -\frac{\partial_r Z_0}{Z_0} \right) = \left( \int_0^{T_i} S_x^2(u) du \right)^{-1/2} B(0,T), i = 1,\cdots,N,$$

$$\frac{\partial}{\partial r} \tilde{d}_i^\pm(\tau, \delta) = \left( \int_0^\tau S_x^2(u) du \right)^{-1/2} \left( -\frac{\partial_r Z_0}{Z_0} \right) = \left( \int_0^\tau S_x^2(u) du \right)^{-1/2} B(0,T), i = 1,\cdots,N \; ([6])$$

to get

$$\partial_r N_{m+1}(d_1^-, \cdots, d_{m+1}^-; A_{m+1}) =$$
$$= B(0,T) \sum_{i=1}^{m+1} \check{N}_{m+1,i}(d_1^-, \cdots, d_{m+1}^-; A_{m+1}) \cdot \left( \int_0^{T_i} S_x^2(u) du \right)^{-1/2} = B(0,T) D_{m+1}^- \geq 0,$$

$$\partial_r N_{m+1}(d_1^+, \cdots, d_m^+, -d_{m+1}^+; A_{m+1}^-) = B(0,T) \left[ \sum_{i=1}^m \check{N}_{m+1,i}(d_1^+, \cdots, d_m^+, -d_{m+1}^+; A_{m+1}^-) \cdot \left( \int_0^{T_i} S_x^2(u) du \right)^{-1/2} \right.$$
$$\left. - \check{N}_{m+1,m+1}(d_1^+, \cdots, d_m^+, -d_{m+1}^+; A_{m+1}^-) \cdot \left( \int_0^{T_{m+1}} S_x^2(u) du \right)^{-1/2} \right] = B(0,T) D_{m+1}^+ ,$$

$$\partial_r N_{m+1}(d_1^-, \cdots, d_m^-, \tilde{d}_{m+1}^-(\tau, \delta); \tilde{A}_{m+1}(\tau)) =$$
$$= B(0,T) \left[ \sum_{i=1}^m \check{N}_{m+1,i}(d_1^-, \cdots, d_m^-, \tilde{d}_{m+1}^-(\tau, \delta); \tilde{A}_{m+1}(\tau)) \cdot \left( \int_0^{T_i} S_x^2(u) du \right)^{-1/2} + \right.$$
$$\left. + \check{N}_{m+1,m+1}(d_1^-, \cdots, d_m^-, \tilde{d}_{m+1}^-(\tau, \delta); \tilde{A}_{m+1}(\tau)) \cdot \left( \int_0^\tau S_x^2(u) du \right)^{-1/2} \right] = B(0,T) \tilde{D}_{m+1}^-(\tau) \geq 0,$$

$$\partial_r N_{m+1}(d_1^+, \cdots, d_m^+, -\tilde{d}_{m+1}^+(\tau, \delta); \tilde{A}_{m+1}^-(\tau)) =$$
$$= B(0,T) \left[ \sum_{i=1}^m \check{N}_{m+1,i}(d_1^+, \cdots, d_m^+, -\tilde{d}_{m+1}^+(\tau, \delta); \tilde{A}_{m+1}^-(\tau)) \cdot \left( \int_0^{T_i} S_x^2(u) du \right)^{-1/2} - \right.$$
$$\left. - \check{N}_{m+1,m+1}(d_1^+, \cdots, d_m^+, -\tilde{d}_{m+1}^+(\tau, \delta); \tilde{A}_{m+1}^-(\tau)) \cdot \left( \int_0^\tau S_x^2(u) du \right)^{-1/2} \right] = B(0,T) \tilde{D}_{m+1}^+(\tau).$$

$$\tag{3.5}$$

Here $\check{N}_{m+1,i}$ is as in (5.8) in the section 5. Using the above notations and estimates, we have

$$\partial_r f_1 = B(0,T) \left\{ F e^{-\sum_{k=0}^{N-1} \lambda_k \Delta T_k} D_N^- + \right.$$
$$\left. + \sum_{m=0}^{N-1} e^{-\sum_{k=0}^{m-1} \lambda_k \Delta T_k} \left[ C_{m+1} e^{-\lambda_m \Delta T_m} D_{m+1}^- + (F + \Sigma_{k=1}^N C_k) \lambda_m e^{-\lambda_m(\tau - T_m)} \int_{T_m}^{T_{m+1}} \tilde{D}_{m+1}^-(\tau) d\tau \right] \right\} = B(0,T) \tilde{f}_1,$$

$$\partial_r f_2 = B(0,T) \sum_{m=0}^{N-1} e^{-\sum_{k=0}^{m-1}(\lambda_k+b)\Delta T_k} \left[ e^{-(\lambda_m+b)\Delta T_m} D_{m+1}^+ + \lambda_m \int_{T_m}^{T_{m+1}} e^{-(\lambda_m+b)(\tau-T_m)} \tilde{D}_{m+1}^+(\tau) d\tau \right] = B(0,T) \tilde{f}_2.$$

Using these notations, we have

$$\partial_r B_0 = \partial_r Z_0 \cdot f_1 + B(0,T)(Z_0 \cdot \tilde{f}_1 + \delta V_0 \cdot \tilde{f}_2).$$

Calculating the duration of our bond, then

$$D = -\frac{\partial_r B_0}{B_0} = \frac{-\partial_r Z_0 \cdot f_1 - B(0,T)(Z_0 \cdot \tilde{f}_1 + \delta V_0 \cdot \tilde{f}_2)}{B_0} = B(0,T) \cdot \left( \frac{Z_0 \cdot f_1}{B_0} - \frac{Z_0 \cdot \tilde{f}_1}{B_0} - \frac{\delta V_0 \cdot \tilde{f}_2}{B_0} \right). \quad (3.6)$$

From $f_1 > 0, f_2 > 0, \tilde{f}_1 > 0$ and (3.4), if $\tilde{f}_2$ is not too small (negative), then $D \leq B(0,T)$, that is, the duration of our bond is smaller than that of the corresponding default free discrete coupon bond. Thus we have proved the following proposition.

**Proposition 1**. If $\tilde{f}_2 \geq \dfrac{-Z_0 \cdot \tilde{f}_1 - \delta V_0 \cdot f_2}{\delta V_0}$, then $D \leq B(0,T)$.

**Remark 8.** If $b = 0$ and $\lambda_k = 0$ ($i = 0,\ldots, N-1$), then the duration (3.6) (considered together with (3.3) and 2.26) of our bond nearly coincides with the result of [1, at page 754] and the only difference comes from the fact that $k$-th coupon is provided as a discounted value of the maturity-value in our model.

## 4. Taxes on the Coupons

In this section we extend the result of the section 2 along the line of the study of [1] on the effect of government taxes that paid on the proceeds of an investment in corporate bonds.

According to [1], State income taxes are only paid on the proceeds of an investment and not on the principal. In this case the payoff to the bond holders is reduced but the equity is not changed. Thus the expected default condition is not changed and default barrier at time $T_i$ is still $K_i$ ($i = 1,\ldots,N$) as calculated in the theorem 1. It means that when the *tax rate* is $\Lambda$ ($> 0$), the *payoff to bondholders* at coupon dates is as follows:

i) At the maturity date $T_N$, $F+(1-\Lambda)C_N$ if $V_{T_N} \geq K_N$ ($=F+C_N$) (firm value is enough large to pay debt principal $F$ and coupon $C_N$); $F+(1-\Lambda)(\delta V_{T_N}-F)$ if $F/\delta \leq V_{T_N} < K_N$ (firm value is enough large to pay debt principal but not enough to pay coupon); $\delta \cdot V_{T_N}$ if $V_{T_N} < F/\delta$ (firm value is not enough large to pay even the principal, let alone the coupon). Here we should note that this structure of the payoff comes from the *implicit* assumption that $F/\delta < F+C_N$ (equally $\delta > (1+c_N)^{-1}$ or $c_N > \delta^{-1} - 1$; we call it the case II) which is possible but generally unlikable because the recovery rate $\delta$ might not be able to be so large provided a coupon rate $c_N = C_N / F$ or the coupon rate $c_N$ might not be able to be so large provided a recovery rate $\delta$. For example, if $\delta = \frac{1}{2}$, then we must have $c_N > 1$ which seems impossible. When $F/\delta \geq F + C_N$ (equally $\delta \leq (1+c_N)^{-1}$ or $c_N \leq \delta^{-1} - 1$; we call it the case I), the payoff to bondholders at the maturity date $T_N$ is $F+(1-\Lambda)C_N$ if $V_{T_N} \geq K_N$ and $\delta \cdot V_{T_N}$ if $V_{T_N} < K_N$. Here we only consider the case I as in [1].

ii) At the *k*-th coupon date $T_i$ ($i = 1,\ldots, N-1$), $(1-\Lambda)C_i \cdot Z(r,T_i;T)$ if $V_{T_i} \geq K_i \cdot Z(r, T_i; T)$; $\delta \cdot V_{T_i}$ if $V_{T_i} < K_i \cdot Z(r, T_i; T)$. (Note that it is possible to consider the case II as at time $T_N$ but we do not consider it since it is generally unlikable.)

Let modify our pricing model (2.18) and (2.19) under consideration of taxes on the coupons provided in the above. We introduce the following notation for simplicity of pricing formulae as the previous subsections.

$$\bar{c}_N = F + (1-\Lambda)C_N;\ \bar{c}_i = (1-\Lambda)C_i, i=1,\cdots,N-1;\ \Phi_i = F + (1-\Lambda)\Sigma_{k=i+1}^{N} C_k. \tag{4.1}$$

That is, $\bar{c}_i$ is the time $T_N$-value of the payoff to bondholders at time $T_i$ ($i=1,\ldots, N$).

Under the above assumption and the notation (4.1), our bond price $\widetilde{B}_i$ satisfies the following PDE in every subinterval $(T_i, T_{i+1})$ ($i = 0, \cdots, N-1$):

$$\frac{\partial \widetilde{B}_i}{\partial t} + \frac{1}{2}\left[s_V^2 V^2 \frac{\partial^2 \widetilde{B}_i}{\partial V^2} + 2\rho s_V s_r V \frac{\partial^2 \widetilde{B}_i}{\partial V \partial r} + s_r^2 \frac{\partial^2 \widetilde{B}_i}{\partial r^2}\right] + (r-b)V \frac{\partial \widetilde{B}_i}{\partial V} + a_r \frac{\partial \widetilde{B}_i}{\partial r} - (r+\lambda_i)\widetilde{B}_i \tag{4.2}$$
$$+ \lambda_i \min\{\delta \cdot V,\ \Phi_i Z(r,t;T)\} = 0,\quad T_i < t < T_{i+1},\ V > 0.$$

If we consider the payoff to bondholders at coupon dates, we can derive the following terminal value conditions:

$$\widetilde{B}_{N-1}(V,r,T_N) = \bar{c}_N \cdot 1\{V \geq K_N\} + \delta V \cdot 1\{V < K_N\},$$
$$\widetilde{B}_i(V,r,T_{i+1}) = [\widetilde{B}_{i+1}(V,r,T_{i+1}) + \bar{c}_{i+1}Z(r,T_{i+1};T)]\cdot 1\{V \geq K_i Z(r,T_{i+1};T)\} + \tag{4.3}$$
$$+ \delta V \cdot 1\{V < K_i Z(r,T_{i+1};T)\},\ V > 0,\quad i = 0,\cdots,N-2.$$

The problem (4.2) and (4.3) with the notation (4.1) is just the *pricing model* of our *defaultable discrete coupon bond* under consideration of *taxes on coupons* and it is the same problem with (2.18) and (2.19). Thus we have the solution representation of it just as in the theorem 2.

**Theorem 3**. *Unless the coupon rates are large relative to* $1/\delta$, *under State tax rate* $\Lambda$, *we have the following representation of the initial price of the our defaultable discrete coupon bond in terms of debt, coupon rates, default recovery rate, default intensity, and initial price of the default free zero coupon bond and initial firm value:*

$$\widetilde{B}_0 = \widetilde{B}_0(F, c_1, \cdots, c_N; \delta, \lambda_0, \cdots, \lambda_{N-1}; Z_0, V_0, \Lambda) =$$
$$= FZ_0\left[e^{-\sum_{k=0}^{N-1}\lambda_k \Delta T_k} N_N(d_1^-, \cdots, d_{N-1}^-, d_N^-; A_N) + \sum_{m=0}^{N-1} e^{-\sum_{k=0}^{m}\lambda_k \Delta T_k}(1-\Lambda)c_{m+1}N_{m+1}(d_1^-, \cdots, d_{m+1}^-; A_{m+1})\right]$$
$$+ \delta V_0 \sum_{m=0}^{N-1} e^{-\sum_{k=0}^{m-1}(\lambda_k+b)\Delta T_k} e^{-(\lambda_m+b)\Delta T_m} N_{m+1}(d_1^+, \cdots, d_m^+, -d_{m+1}^+; A_{m+1}^-) +$$
$$+ \sum_{m=0}^{N-1} \lambda_m \int_{T_m}^{T_{m+1}}\left[\delta V_0 e^{-\sum_{k=0}^{m-1}(\lambda_k+b)\Delta T_k} e^{-(\lambda_m+b)(\tau-T_m)} N_{m+1}(d_1^+, \cdots, d_m^+, -\hat{d}_{m+1}^+(\tau,\delta,\Lambda); \widetilde{A}_{m+1}^-(\tau)) +\right.$$
$$\left. + FZ_0[1 + \Sigma_{k=1}^{N}(1-\Lambda)c_k]e^{-\sum_{k=0}^{m-1}\lambda_k \Delta T_k} e^{-\lambda_m(\tau-T_m)} N_{m+1}(d_1^-, \cdots, d_m^-, \hat{d}_{m+1}^-(\tau,\delta,\Lambda); \widetilde{A}_{m+1}^-(\tau))\right]d\tau. \tag{4.4}$$

Here $Z_0, N_m(a_1,\cdots,a_m ; A), A_m, A_m^-, \tilde{A}_m(\tau)$ and $\tilde{A}_m^-(\tau)$ are the same as in the theorem 2 and $d_i^\pm$, $\hat{d}_i^\pm(\tau,\delta,\Lambda)$ are given by

$$d_i^\pm = \left(\int_0^{T_i} S_x^2(u)du\right)^{-1/2}\left(\ln\frac{V_0}{Z_0 K_i} - bT_i \pm \frac{1}{2}\int_0^{T_i} S_x^2(u)du\right), \ i=\overline{1,N-1};$$

$$d_N^\pm = \left(\int_0^{T_N} S_x^2(u)du\right)^{-1/2}\left(\ln\frac{V_0}{FZ_0(1+c_N)} - bT_N \pm \frac{1}{2}\int_0^{T_N} S_x^2(u)du\right);$$

$$\hat{d}_i^\pm(\tau,\delta,\Lambda) = \left(\int_0^\tau S_x^2(u)du\right)^{-1/2}\left(\ln\frac{\delta V_0}{FZ_0[1+\Sigma_{k=i}^N(1-\Lambda)c_k]} - b\tau \pm \frac{1}{2}\int_0^\tau S_x^2(u)du\right), T_{i-1}\leq\tau<T_i; i=\overline{1,N}.$$

**Remark 9.** The financial meaning of every term of the formula (4.4) is similar with the theorem 2. Note that If $b = 0$ and $\lambda_k = 0$ ($i = 0,\ldots, N–1$), then our pricing formula (4.4) nearly coincides with the formula (10) of [1, at page 756] and the only difference comes from the fact that $k$-th coupon is provided as a discounted value of the maturity-value in our model.

**Remark 10.** As in the section 2, the equation of the relative price $\tilde{u}_i(x,t) = \tilde{B}_i / Z$ of the bond is an inhomogenous Black-Scholes equation with discontinuous terminal value. Thus using the results of [9], we can investigate such properties of $u_i(x,t)$ as monotonicity, boundedness or gradient estimate and so on. For example, if we denote $\alpha = 1 - \Lambda c_N/(1+c_N)$, then $\bar{c}_N = \alpha K_N$. If $\alpha \geq \delta$, then the terminal value $\tilde{u}_{N-1}(x,T_N)$ is increasing and thus by the theorem 4 of [9], $\tilde{u}_{N-1}(x,t)(t<T_N)$ is $x$- increasing. If $\alpha < \delta$, then the terminal value $\tilde{u}_{N-1}(x,T_N)$ is not increasing and thus we cannot say that $\tilde{u}_{N-1}(x,t)(t<T_N)$ is $x$ – increasing (but in this case the default recovery can be greater than the normal payoff to bondholders and this seems *not appropriate* in financial meaning). (See the figure 1.)

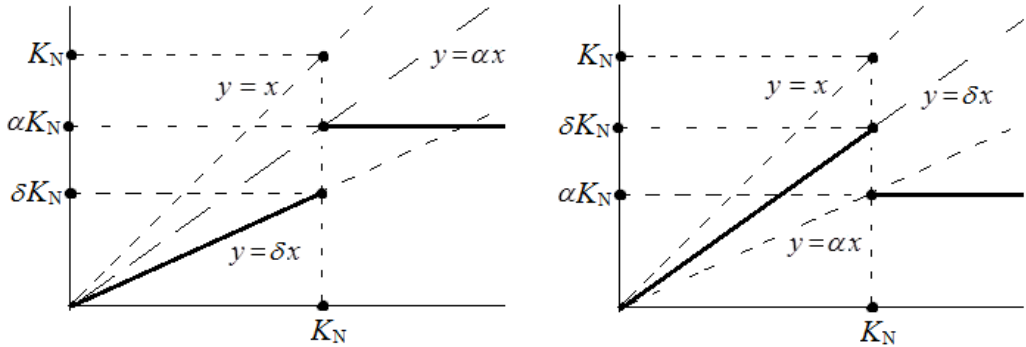

Figure 1. The graph of $\tilde{u}_{N-1}(x,T_N)$. (a) $\alpha \geq \delta$, (b) $\alpha < \delta$

## 5. Appendix

### The Proof of Theorem 1

Now we solve the problem (2.10) and (2.11). If we use the *change of numeraire*

$$x = V/Z(r,t;T), \ e_i(x,t) = E_i(V, r, t)/Z(r,t;T), \ T_i < t \leq T_{i+1}, \quad (5.1)$$

then we have the following problems:

Comprehensive Unified Models of Structural and Reduced form models for Defaultable Fixed Income Bonds
(Part I: One factor models, Part II: Two factors-Model)

$$\frac{\partial e_i}{\partial t} + \frac{1}{2} S_x^2(t) \frac{\partial^2 e_i}{\partial x^2} - bx \frac{\partial e_i}{\partial x} + a_r \frac{\partial E_i}{\partial r} - \lambda_i e_i = 0, \; T_i < t < T_{i+1}, \; x > 0, \tag{5.2}$$

$$e_{N-1}(x, T_N) = x \cdot 1\{V \geq K_N\} - \bar{c}_N \cdot 1\{V \geq K_N\},$$
$$e_i(x, T_{i+1}) = [e_{i+1}(x, T_{i+1}) - \bar{c}_{i+1}] \cdot 1\{e_{i+1}(x, T_{i+1}) \geq \bar{c}_{i+1}\}, \; i = 0, \cdots, N-2. \tag{5.3}$$

The equation (5.2) is the Black-Scholes equation with the short rate $\lambda_i$, the dividend rate $\lambda_i + b$ and the volatility $S_x(t)$.

When $i = N-1$, the problem (5.2) and (5.3) is a generalized European option problem in the meaning of [2] and its solution is given by the binary options as follows:

$$e_{N-1}(x,t) = A^+_{K_N}(x,t; T_N; \lambda_{N-1}, \lambda_{N-1}+b, S_x(\cdot)) - \bar{c}_N B^+_{K_N}(x,t; T_N; \lambda_{N-1}, \lambda_{N-1}+b, S_x(\cdot)), \tag{5.4}$$
$$T_{N-1} < t \leq T_N, \; x > 0.$$

Here $A^+_{K_N}(x,t; T_N; \lambda_{N-1}, \lambda_{N-1}+b, S_x(\cdot))$, $B^+_{K_N}(x,t; T_N; \lambda_{N-1}, \lambda_{N-1}+b, S_x(\cdot))$ are the prices of the asset and bond binary options with the coefficients $\lambda_{N-1}$, the dividend rate $\lambda_{N-1}+b$ and the volatility $S_x(t)$ (see the formulae (2.9) in [10]). For simplicity of notation, we use the relation (2.12) of [10] to rewrite (5.4) as the binary options with 0-short rate, $b$-dividend rate and the volatility $S_x(t)$:

$$e_{N-1}(x,t) = e^{-\lambda_{N-1}(T_N - t)}[A^+_{K_N}(x,t; T_N; 0, b, S_x(\cdot)) - \bar{c}_N B^+_{K_N}(x,t; T_N; 0, b, S_x(\cdot))],$$
$$T_{N-1} < t \leq T_N, \; x > 0.$$

For the study of the next step, we investigate the properties of $e_{N-1}(x,t)$. The terminal value $e_{N-1}(x, T_N) = f(x)$ is downward convex and $\min_x f'(x) = 0$, $\max_x f'(x) = 1$. From the theorem 2 and theorem 3 of [9], $e_{N-1}(x,t)$ is $x$-increasing and $x$-downward convex, in particular, we have

$$0 < \frac{\partial e_{N-1}}{\partial x}(x, T_{N-1}) < e^{-(\lambda_{N-1}+b)(T_N - T_{N-1})} \leq 1, \; x > 0. \tag{5.5}$$

Thus the equation $e_{N-1}(x, T_{N-1}) = \bar{c}_{N-1}$ has unique root $K_{N-1}$ and $e_{N-1}(x, T_{N-1}) \geq \bar{c}_{N-1} \Leftrightarrow x \geq K_{N-1}$. (Note that $\bar{c}_{N-1} = 0 \Leftrightarrow K_{N-1} = 0$.)

Now consider the case when $i = N-2$. In this case, the equation (5.2) is the Black-Scholes equation with the short rate $\lambda_{N-2}$, the dividend rate $\lambda_{N-2}+b$ and the volatility $S_x(t)$, and the terminal value condition (5.3) can be written as

$$e_{N-2}(x, T_{N-1}) = [e_{N-1}(x, T_{N-1}) - \bar{c}_{N-1}] \cdot 1\{x \geq K_{N-1}\} =$$
$$= [A^+_{K_N}(x, T_{N-1}; T_N; \lambda_{N-1}, \lambda_{N-1}+b, S_x(\cdot)) \cdot 1\{x \geq K_{N-1}\}$$
$$- \bar{c}_N B^+_{K_N}(x, T_{N-1}; T_N; \lambda_{N-1}, \lambda_{N-1}+b, S_x(\cdot)) \cdot 1\{x \geq K_{N-1}\} - \bar{c}_{N-1} \cdot 1\{x \geq K_{N-1}\}.$$

Here we use the relation (2.12) of [10] to get

$$e_{N-2}(x, T_{N-1}) = e^{-(\lambda_{N-1} - \lambda_{N-2})(T_N - T_{N-1})}[A^+_{K_N}(x, T_{N-1}; T_N; \lambda_{N-2}, \lambda_{N-2}+b, S_x(\cdot)) \cdot 1\{x \geq K_{N-1}\}$$
$$- \bar{c}_N B^+_{K_N}(x, T_{N-1}; T_N; \lambda_{N-1}, \lambda_{N-1}+b, S_x(\cdot)) \cdot 1\{x \geq K_{N-1}\}] - \bar{c}_{N-1} \cdot 1\{x \geq K_{N-1}\}.$$

This is a linear combination of the terminal values of second order binary options in the meaning of [2] and the solution $e_{N-2}(x,t)$ is given by the second order binary options:

$$e_{N-2}(x,t) = e^{-(\lambda_{N-1}-\lambda_{N-2})\Delta T_{N-1}} [A^{++}_{K_{N-1}K_N}(x,t;T_{N-1},T_N;\lambda_{N-2},\lambda_{N-2}+b,S_x(\cdot))$$
$$-\bar{c}_N B^{++}_{K_{N-1}K_N}(x,t;T_{N-1},T_N;\lambda_{N-2},\lambda_{N-2}+b,S_x(\cdot))] \quad (5.6)$$
$$-\bar{c}_{N-1} B^+_{K_{N-1}}(x,t;T_{N-1};\lambda_{N-2},\lambda_{N-2}+b,S_x(\cdot)), \quad T_{N-2} < t \le T_{N-1}, \ x > 0.$$

Here $A^{++}_{K_{N-1}K_N}(x,t;T_{N-1},T_N;\lambda_{N-2},\lambda_{N-2}+b,S_x(\cdot))$, $B^{++}_{K_{N-1}K_N}(x,t;T_{N-1},T_N;\lambda_{N-2},\lambda_{N-2}+b,S_x(\cdot))$ are the prices of the second order asset and bond binary options with the coefficients $\lambda_{N-2}$, the dividend rate $\lambda_{N-2}+b$ and the volatility $S_x(t)$ (see the formulae (2.10) of [10]). For simplicity of notation, we use the relation (2.12) of [10] to rewrite (5.4) as the binary options with 0-short rate, $b$-dividend rate and the volatility $S_x(t)$:

$$e_{N-2}(x,t) = e^{-\lambda_{N-2}(T_{N-1}-t)} \{ e^{-\lambda_{N-1}\Delta T_{N-1}} [A^{++}_{K_{N-1}K_N}(x,t;T_{N-1},T_N;0,b,S_x(\cdot))$$
$$-\bar{c}_N B^{++}_{K_{N-1}K_N}(x,t;T_{N-1},T_N;0,b,S_x(\cdot))]$$
$$-\bar{c}_{N-1} B^+_{K_{N-1}}(x,t;T_{N-1};0,b,S_x(\cdot)) \}, \quad T_{N-2} < t \le T_{N-1}, \ x > 0.$$

For the study of the next step, we investigate the properties of $e_{N-2}(x,t)$. From the properties of $e_{N-1}(x,T_{N-1})$, the terminal value $e_{N-2}(x,T_{N-1}) = [e_{N-1}(x,T_{N-1})-\bar{c}_{N-1}] \cdot 1\{x \ge K_{N-1}\} = f(x)$ is downward convex and $\min_x f'(x) = 0$, $\max_x f'(x) \le e^{-(\lambda_{N-1}+b)\Delta T_{N-1}}$. From the theorem 2 and theorem 3 of [9], $e_{N-2}(x,t)$ is $x$-increasing and $x$-downward convex, in particular,

$$0 < \frac{\partial e_{N-2}}{\partial x}(x,T_{N-2}) < e^{-(\lambda_{N-2}+b)\Delta T_{N-2} - (\lambda_{N-1}+b)\Delta T_{N-1}} \le 1, \ x > 0. \quad (5.7)$$

Thus the equation $e_{N-2}(x,T_{N-2}) = \bar{c}_{N-2}$ has unique root $K_{N-2}$ and $e_{N-2}(x,T_{N-2}) \ge \bar{c}_{N-2} \Leftrightarrow x \ge K_{N-2}$. (Note that $\bar{c}_{N-2} = 0 \Leftrightarrow K_{N-2} = 0$.)

Thus when $i = N-3$, the terminal value (5.3) can be written as

$$e_{N-3}(x,T_{N-2}) = [e_{N-2}(x,T_{N-2})-\bar{c}_{N-2}] \cdot 1\{x \ge K_{N-2}\} =$$
$$= e^{-(\lambda_{N-1}-\lambda_{N-2})\Delta T_{N-1}} [A^{++}_{K_{N-1}K_N}(x,T_{N-2};T_{N-1},T_N;\lambda_{N-2},\lambda_{N-2}+b,S_x(\cdot)) \cdot 1\{x \ge K_{N-2}\}$$
$$-\bar{c}_N B^{++}_{K_{N-1}K_N}(x,T_{N-2};T_{N-1},T_N;\lambda_{N-2},\lambda_{N-2}+b,S_x(\cdot)) \cdot 1\{x \ge K_{N-2}\}]$$
$$-\bar{c}_{N-1} B^+_{K_{N-1}}(x,T_{N-2};T_{N-1};\lambda_{N-2},\lambda_{N-2}+b,S_x(\cdot)) \cdot 1\{x \ge K_{N-2}\} - \bar{c}_{N-2} \cdot 1\{x \ge K_{N-2}\}.$$

But in this case, the equation (5.2) is the Black-Scholes equation with the short rate $\lambda_{N-3}$, the dividend rate $\lambda_{N-3}+b$ and the volatility $S_x(t)$ and thus we use the relation (2.12) of [10] to get

$$e_{N-3}(x,T_{N-2}) =$$
$$= e^{-\lambda_{N-1}\Delta T_{N-1} - \lambda_{N-2}\Delta T_{N-2} + \lambda_{N-3}(T_N - T_{N-2})} [A^{++}_{K_{N-1}K_N}(x,T_{N-2};T_{N-1},T_N;\lambda_{N-3},\lambda_{N-3}+b,S_x(\cdot)) \cdot 1\{x \ge K_{N-2}\}$$
$$-\bar{c}_N B^{++}_{K_{N-1}K_N}(x,T_{N-2};T_{N-1},T_N;\lambda_{N-3},\lambda_{N-3}+b,S_x(\cdot)) \cdot 1\{x \ge K_{N-2}\}] - \bar{c}_{N-2} \cdot 1\{x \ge K_{N-2}\}$$
$$-\bar{c}_{N-1} e^{-\lambda_{N-2}\Delta T_{N-2} + \lambda_{N-3}\Delta T_{N-2}} B^+_{K_{N-1}}(x,T_{N-2};T_{N-1};\lambda_{N-3},\lambda_{N-3}+b,S_x(\cdot)) \cdot 1\{x \ge K_{N-2}\}.$$

This is a linear combination of the terminal values of third order binary options in the meaning of [9] and the solution $e_{N-3}(x,t)$ is given by the third order binary options:

$$e_{N-3}(x,t) = e^{-\lambda_{N-1}\Delta T_{N-1} - \lambda_{N-2}\Delta T_{N-2} + \lambda_{N-3}(T_N - T_{N-2})}[A^{+\ +\ +}_{K_{N-2}K_{N-1}K_N}(x,t;T_{N-2},T_{N-1},T_N;\lambda_{N-3},\lambda_{N-3}+b,S_x(\cdot))$$
$$- \bar{c}_N B^{+\ +\ +}_{K_{N-2}K_{N-1}K_N}(x,t;T_{N-2},T_{N-1},T_N;\lambda_{N-3},\lambda_{N-3}+b,S_x(\cdot))]$$
$$- \bar{c}_{N-1} e^{-\lambda_{N-2}\Delta T_{N-2} + \lambda_{N-3}\Delta T_{N-2}} B^{+\ +}_{K_{N-2}K_{N-1}}(x,t;T_{N-2},T_{N-1};\lambda_{N-3},\lambda_{N-3}+b,S_x(\cdot))$$
$$- \bar{c}_{N-2} B^+_{K_{N-2}}(x,t;T_{N-2};\lambda_{N-3},\lambda_{N-3}+b,S_x(\cdot)), \quad T_{N-3} < t \le T_{N-2}, \ x > 0.$$

Here $A^{+\ +\ +}_{K_{N-2}K_{N-1}K_N}, B^{+\ +\ +}_{K_{N-2}K_{N-1}K_N}$ are the prices of the third order asset and bond binary options with the coefficients $\lambda_{N-3}$, the dividend rate $\lambda_{N-3}+b$ and the volatility $S_x(t)$ (see the formulae (2.11) of [10]). For simplicity of notation, we use the relation (2.11) of [10] to rewrite it as the binary options with 0-short rate, $b$-dividend rate and the volatility $S_x(t)$:

$$e_{N-3}(x,t) = e^{-\lambda_{N-3}(T_{N-2}-t)}\{e^{-\lambda_{N-1}\Delta T_{N-1}-\lambda_{N-2}\Delta T_{N-2}} A^{+\ +\ +}_{K_{N-2}K_{N-1}K_N}(x,t;T_{N-2},T_{N-1},T_N;0,b,S_x(\cdot))$$
$$- \bar{c}_N e^{-\lambda_{N-1}\Delta T_{N-1}-\lambda_{N-2}\Delta T_{N-2}} B^{+\ +\ +}_{K_{N-2}K_{N-1}K_N}(x,t;T_{N-2},T_{N-1},T_N;0,b,S_x(\cdot))$$
$$- \bar{c}_{N-1} e^{-\lambda_{N-2}\Delta T_{N-2}} B^{+\ +}_{K_{N-2}K_{N-1}}(x,t;T_{N-2},T_{N-1};0,b,S_x(\cdot))$$
$$- \bar{c}_{N-2} B^+_{K_{N-2}}(x,t;T_{N-2};0,b,S_x(\cdot))\}, \quad T_{N-3} < t \le T_{N-2}, \ x > 0.$$

By induction the theorem 1 is proved. (QED)

In the calculation *of duration for defaultable bond, we use the* derivatives of multi-variate normal distribution functions. Let

$$N_m(a_1(x),\cdots,a_m(x);A) = \int_{-\infty}^{a_1(x)} \cdots \int_{-\infty}^{a_m(x)} \frac{1}{(\sqrt{2\pi})^m}\sqrt{\det A}\exp(-\frac{1}{2}y^\perp Ay)dy.$$

**Lemma 1**. *The derivatives of multi-variate normal distribution functions are provided as follows:*

$$\partial_x N_m(a_1(x),\cdots,a_m(x);A) = \sum_{i=1}^m \breve{N}_{m,i}(a_1(x),\cdots,a_m(x);A)a_i'(x). \tag{5.8}$$

*Here we used the following notations:*

$$\breve{N}_{m,i}(a_1(x),\cdots,a_m(x);A) = \int_{-\infty}^{a_1(x)}\cdots\int_{-\infty}^{a_{i-1}(x)}\int_{-\infty}^{a_{i+1}(x)}\cdots\int_{-\infty}^{a_m(x)} \frac{\sqrt{\det A}}{(\sqrt{2\pi})^m}\exp\left(-\frac{1}{2}\tilde{y}_i(x)^\perp A\,\tilde{y}_i(x)\right)d\breve{y}_i,$$

$$\tilde{y}_i(x)^\perp = (y_1,\cdots,y_{i-1},a_i(x),y_{i+1},\cdots,y_m), \ d\breve{y}_i = dy_1\cdot dy_{i-1}dy_{i+1}\cdots dy_m \ ; i=1,\cdots,m. \tag{5.9}$$

**Proof**: From the lemma on differentiation of integral with parameter we have

$$\partial_x N_m(a_1(x),\cdots,a_m(x);A) = \int_{-\infty}^{a_1(x)} \frac{\partial}{\partial x}\int_{-\infty}^{a_2(x)}\cdots\int_{-\infty}^{a_m(x)} \frac{1}{(\sqrt{2\pi})^m}\sqrt{\det A}\exp(-\frac{1}{2}y^\perp Ay)dy +$$
$$+ \int_{-\infty}^{a_2(x)}\cdots\int_{-\infty}^{a_m(x)} \frac{1}{(\sqrt{2\pi})^m}\sqrt{\det A}\exp\left(-\frac{1}{2}y^\perp Ay\right)\bigg|_{y_1 = a_1(x)} d\breve{y}_1 \cdot a_1'(x) =$$
$$= \breve{N}_{m,1}(a_1(x),\cdots,a_m(x);A)a_1'(x) + \int_{-\infty}^{a_1(x)}\frac{\partial}{\partial x}\int_{-\infty}^{a_2(x)}\cdots\int_{-\infty}^{a_m(x)} \frac{1}{(\sqrt{2\pi})^m}\sqrt{\det A}\exp(-\frac{1}{2}y^\perp Ay)dy.$$

Going on using the lemma on differentiation of integral with parameter in the second term in the above expression, we can prove (5.8) by induction. (QED)